\documentclass{aa}
\usepackage{txfonts}
\usepackage{graphicx}	
\usepackage{amsmath}	
\usepackage{amssymb}	
\usepackage{caption,subcaption}
\usepackage{textgreek}
\usepackage{verbatim}
\usepackage{gensymb}
\usepackage{xcolor}

\usepackage{ulem}
\usepackage[export]{adjustbox}
\usepackage[colorlinks=true, citecolor=blue]{hyperref}

\begin{document}

   \title{The cosmic radio dipole: Bayesian estimators on new and old radio surveys}

   \author{J.~D. Wagenveld\inst{1}
          \and H-R. Kl\"{o}ckner\inst{1}
          \and D.~J. Schwarz\inst{2}
          }

   \institute{Max-Planck Institut fur Radioastronomie, 
              Auf dem H\"{u}gel 69, 
              53121 Bonn, Germany
        \and Fakult\"{a}t f\"{u}r Physik, 
             Universit\"{a}t Bielefeld, 
             Postfach 100131, 
             33501 Bielefeld, Germany
            }


 
    \abstract{The cosmic radio dipole is an anisotropy in the number counts of radio sources, analogous to the dipole seen in the cosmic microwave background (CMB). Measurements of source counts of large radio surveys have shown that though the radio dipole is generally consistent in direction with the CMB dipole, the amplitudes are in tension. These observations present an intriguing puzzle as to the cause of this discrepancy, with a true anisotropy breaking with the assumptions of the cosmological principle, invalidating the most common cosmological models that are built on these assumptions. We present a novel set of Bayesian estimators to determine the cosmic radio dipole and compare the results with commonly used methods on the Rapid ASKAP Continuum Survey (RACS) and the NRAO VLA Sky Survey (NVSS) radio surveys. In addition, we adapt the Bayesian estimators to take into account systematic effects known to affect such large radio surveys, folding information such as the local noise floor or array configuration directly into the parameter estimation. The enhancement of these estimators allows us to greatly increase the amount of sources used in the parameter estimation, yielding tighter constraints on the cosmic radio dipole estimation than previously achieved with NVSS and RACS. We extend the estimators further to work on multiple catalogues simultaneously, leading to a combined parameter estimation using both NVSS and RACS. The result is a dipole estimate that perfectly aligns with the CMB dipole in terms of direction but with an amplitude that is three times as large, and a significance of $4.8\sigma$. This new dipole measurement is made to an unprecedented level of precision for radio sources, which is only matched by recent results using infrared quasars.}


\maketitle

\section{Introduction}

The cosmic microwave background (CMB) and the structures seen therein are a signpost for large scale structure in the universe. Besides the well-studied low level ($\Delta T/T \sim 10^{-5}$) anisotropies, the largest anisotropy seen in the CMB presents itself in the cosmic dipole ($\Delta T/T \sim 10^{-3}$), an effect attributed to the movement of the Solar system with respect to the CMB restframe. The velocity of the solar system as derived from the amplitude of the CMB dipole is $v = 369.82\pm0.11\ \mathrm{km\ s^{-1}}$ \citep{Aghanim2020}, assuming that the dipole is entirely caused by the observers' motion w.r.t. the CMB. 

\citet{Ellis1984} first proposed measuring the dipole in the number counts of radio sources, as the radio population outside of the Galactic plane mostly consists of extragalactic sources that are expected to be part of and trace large scale structure. The first significant measurement of the cosmic radio dipole\footnote{Both `cosmic radio dipole' or `radio dipole' are used in this paper to refer to the dipole observed in the number counts of radio sources.} was reported by \citet{Blake2002} using the National Radio Astronomy Observatory (NRAO) Very Large Array (VLA) Sky Survey \citep[NVSS,][]{Condon1998}, which agreed with the CMB dipole within uncertainties. However, subsequent studies using the NVSS found the amplitude of the cosmic radio dipole  to be significantly larger than that of the CMB dipole. \citet{Singal2011} found a radio dipole  to be four times larger, which was corroborated by \citet{Rubart2013}, in both cases with a $3\sigma$ significance. Apart from to the NVSS measurement, the cosmic radio dipole  was measured with other radio surveys, such as the Westerbork Northern Sky Survey \citep[WENSS,][]{Rengelink1997}, the Tata Institute for Fundamental Research (TIFR) Giant Metrewave Radio Telescope (GMRT) Sky Surveys first alternative data release \citep[TGSS ADR1,][]{Intema2017}, and the Sydney University Molonglo Sky Survey \citep[SUMSS,][]{Mauch2003}. Though depending on the survey and employed estimator, the amplitude of the radio dipole is consistently larger than the amplitude of the CMB dipole \citep[see][for an overview]{Siewert2021}, while the direction of the radio dipole remains consistent with that of the CMB, albeit with considerable uncertainty.

Most significant are the results from \citet{Secrest2021} and \citet{Secrest2022}, which find a dipole amplitude over twice that of the CMB at a significance of $4.9\sigma$ using Wide-field Infrared Survey Explorer \citep[WISE,][]{Wright2010} measurements of quasars. \citet{Secrest2022} perform a joint analysis of NVSS radio galaxies and WISE quasars, with a resulting significance of $5.1\sigma$. A Bayesian estimator based on Poisson statistics was recently utilised by \citet{Dam2022} for the first time to measure the cosmic radio dipole with the WISE quasar sample from \citet{Secrest2021}, yielding a dipole amplitude 2.7 times larger than the CMB dipole with a significance of $5.7\sigma$. Ultimately, a discrepancy between a dipole in the CMB and other observables points to an unknown effect to the data, which is increasingly unlikely to be systematic among all the different probes, as the NVSS and WISE samples for example are independent both in terms of source population as well as systematic effects. This points to an unexpected anisotropy in the large scale structure of the Universe, something which breaks with the core assumptions of the Cosmological principle. If true, this poses a major problem for cosmologies that are based on the Friedmann–Lemaître–Robertson–Walker metric, such as \textLambda-CDM.

One problem that remains persistent even with ever larger datasets is a lack of homogeneity caused by systematic effects to the data. Systematic effects that are unaccounted for can greatly bias dipole estimates, such that conservative cuts in the data must be made to eliminate biases as much as possible. Already in the first measurement of the cosmic radio dipole  using the NVSS \citep{Blake2002}, a persistent systematic effect was identified causing large differences in source density as a function of declination, inducing an artificial north-south anisotropy in the data. To eliminate this effect and avoid biasing dipole estimates, conservative cuts in flux density have to be made in such catalogues, which greatly reduces the number of usable sources. 

In \citet{Wagenveld2023}, a deep analysis of ten pointings from the MeerKAT Absorption Line Survey \citep{Gupta2016} was presented, with a focus on mitigating biases that could affect a measurement of the cosmic radio dipole. Being able to account for systematic effects allows less strict flux density cuts, increasing homogeneity of the catalogues and number of sources that can be used for a dipole estimate. This approach was made possible by having direct access to meta data and data products within the processing steps of the survey from calibration to imaging and source finding, which is commonly not the case in dipole studies. In this work we approach this problem from the outside, and show how dipole estimates can be improved with the information present in modern radio catalogues. We present new dipole estimators, constructing likelihoods for estimating dipole parameters that can take this information into account, as well as an estimator that combines different catalogues for an improved dipole estimate. Given the proper information, these estimators are able to account for systematic effects on number counts in radio surveys and remove the need to cut large amounts of data.

This paper is organised as follows. In Section~\ref{sec:source_counts} we describe the statistics of radio source counts and the dipole effect. In Section~\ref{sec:estimators} we introduce the estimators that will be used to infer the radio dipole  parameters, using the data sets described in Section~\ref{sec:data}. The results obtained are given in Section~\ref{sec:results}. The implications and caveats of the results are discussed in Section~\ref{sec:discussion}. In Section~\ref{sec:conclusion} we summarise the findings of this paper.

\section{Radio source counts and the dipole}
\label{sec:source_counts}

The majority of bright sources at radio wavelengths outside of the Galactic plane are active galactic nuclei (AGN) and have a redshift distribution that peaks at $z\sim0.8$ \citep[e.g.][]{Condon2016}. As such, radio sources are expected to trace the background, and should thus comply with isotropy and homegeneity on the largest scales. Following the cosmological principle, the surface density of sources should therefore be independent of location on the sky. The naive expectation is a distribution of radio sources that are independent, identical, and point-like, which defines a Poisson point process. By discretising the sky into regions of finite size, the number of sources per region will follow a Poisson distribution. The probability density distribution
\begin{equation}
    p(n) = \frac{\lambda^ne^{-\lambda}}{n!},
    \label{eq:poisson}
\end{equation}
is entirely parametrised by the variable $\lambda$, which describes both the mean and variance of the distribution. In actual radio data, some deviations from a perfect Poisson distribution are expected due to clustering and the presence of sources with multiple components. The severity of these effects depend largely on things such as survey depth, angular resolution, and observing frequency, and thus are difficult to assess without a thorough analysis of the survey. Such an analysis is beyond the scope of this work, but would follow a structure similar to that of \citet{Siewert2020}, who demonstrate for the LOFAR Two-Metre Sky Survey first data release \citep[LoTSS DR1,][]{Shimwell2019} that the distribution of source counts converges to a Poisson distribution if applying more strict flux density cuts. For the analysis presented in this work, we assume that the effects of clustering and multi-component sources are negligible on a dipole estimate. 

The spectral features and number count relations of typical radio sources make them uniquely suitable for a dipole measurement. For most sources the dominant emission mechanism at radio wavelengths is synchrotron radiation, the spectral behaviour of which is well described by a power law, 
\begin{equation}
    S\propto\nu^{-\alpha},
\end{equation}
with a characteristic spectral index $\alpha$. For synchrotron emission the typical value of $\alpha$ is around 0.75 \citep[e.g.][]{Condon1992}, and this value has been assumed for most dipole studies at radio wavelengths \citep{Ellis1984,Rubart2013,Siewert2021}. Furthermore, the number density of radio sources follows a power-law relation with respect to the flux density $S$ above which the counts are taken,
\begin{equation}
    \frac{\mathrm{d}N}{\mathrm{d}\Omega} (>S) \propto S^{-x}.
\end{equation}
The value of $x$ can differ per survey, depending on choice of flux density cut and frequency of the catalogue, but usually takes values of 0.75-1.0.  

In the frame of the moving observer, given a velocity $\beta=v/c$, a systemic Doppler effect shifts the spectra of sources, which affects the flux density of these sources. Additionally sources are Doppler boosted from the point of view of the moving observer, which further affects the observed flux density of sources. Depending on their angular distance from the direction of motion $\theta$, the flux densities are shifted by
\begin{equation}
    S_{obs} = (1+\beta\cos\theta)^{1+\alpha} S_{rest}.
    \label{eq:dipole_flux}
\end{equation}
Thus given a flux limited survey of radio sources, more sources appear above the minimum observable flux in the direction of the motion, and less will appear in the opposite direction. Finally, relativistic aberration caused by motion of the observer shifts the positions of sources \textit{towards} the direction of motion, causing a further increase in number counts in the direction of motion, 
\begin{align}
    \tan\theta_{obs} = \frac{\sin\theta_{rest}}{\beta - \cos\theta_{rest}}.
    \label{eq:dipole_pos}
\end{align}
As the fluxes and positions of the sources are shifted, we observe the dipole as an asymmetry in the number counts of radio sources. Combining these effects to first order in $\beta$, thus assuming that $v \ll c$, shows the expected dipole amplitude for a given survey
\begin{align}
    \vec{d} &= \mathcal{D}\cos\vec{\theta}, \\
    \mathcal{D} &= [2 + x(1+\alpha)]\beta.
    \label{eq:dipole}
\end{align}
As such, we can directly infer the velocity of the observer by measuring the dipole effect on the number counts of sources. However, this necessitates the assumption that dipole is entirely caused by the motion of the observer. Given the observed discrepancy between the CMB dipole and the radio dipole , it would be equally appropriate to assume that part of the observed radio dipole is caused by a different (and as of yet unknown) effect. 

Given a dipole characterised by Equation~(\ref{eq:dipole}), different dipole amplitudes $\mathcal{D}$ are expected to be seen depending on the data set being used. Though aberration always has an equal effect, the effect of Doppler shift is determined by spectral index $\alpha$ of the sources, and both this and the Doppler boosting effect depends on the flux distribution of sources, characterised by power-law index $x$. Most often single values are assumed for these quantities, from which the expectation of the dipole amplitude can be derived. These quantities will differ at least between different surveys, so we will derive them for each survey separately. Although the entire flux distribution of any survey can hardly be characterised with a single power-law relation, the dipole in number counts will be caused by sources near the flux density threshold, making the power-law fit near this threshold the most appropriate choice for deriving $x$. For the entire range of frequencies we are considering in this work, we assume a spectral index $\alpha=0.75$, considering the synchrotron emission of radio sources which dominates the spectrum of radio sources below 30~GHz \citep{Condon1992}.

\section{Dipole estimators}
\label{sec:estimators}

For measuring the cosmic radio dipole  different types of estimators have been used that for the most part yield consistent results. Most commonly used are linear and quadratic estimators \citep[e.g.][]{Singal2011,Rubart2013,Siewert2021}. Linear estimators essentially sum up all source positions, and thus, by design, will point towards the largest anisotropy in the data. However, the recovered amplitude from linear estimators is inherently biased. Furthermore, because of its sensitivity to anisotropies, any gaps or systematic effects in the data can introduce biases in the estimate of the dipole direction. To not bias the estimator w.r.t.\ dipole direction, a mask must be created such that the map remains point symmetric w.r.t.\ the observer, or a `masking correction' must be applied \citep[e.g.][]{Singal2011,Rubart2013}. Consequently, missing data features such as the Galactic plane must be mirrored to maintain symmetry, removing even more data. The quadratic estimator compares expected number counts with a model, providing a chi-squared test of the data with respect to a model of the dipole \citep[e.g.][]{Siewert2021}. The best-fit dipole parameters are then retrieved by minimising $\chi^2$. Though the cost is imposing a dipole model on the data, the estimate is not biased by the spatial gaps in the data that are ubiquitous in radio surveys. 

Both aforementioned estimators are sensitive to anisotropies in the data introduced by systematic effects. Most commonly, these systematics affect the sensitivity of the survey in different parts of the sky, making the most straightforward solution to cut out all sources below some flux density. Given this assessment however, we might expect that information of the sensitivity of the survey in different parts of the sky can help alleviate these biases. The new generation of radio surveys boasts catalogues with a wealth of information, including the local root-mean-square (RMS) noise, which we will exploit in this work. 

To improve sensitivity to the dipole several attempts have been made to combine different radio catalogues. Both \citet{Colin2017} and \citet{Darling2022} worked on combining different radio surveys, using different techniques to deal with systematic differences between the catalogues. Here we will provide an alternative method to combine catalogues for increased sensitivity to the dipole, while accounting for systematic differences between the catalogues. 

\subsection{Quadratic estimator}
\label{sec:quadratic_estimator}

To control for differences in pixelation and masking strategies between this and previous works, introducing a new estimator warrants a comparison with known methods of dipole estimation. The quadratic estimator is the closest analogy to the Poisson estimator which we use to produce our main results, in that it is insensitive to gaps in the data. It's effectiveness and results on multiple large radio surveys are presented in \citet{Siewert2021}. The estimator is based on the Pearson's chi-squared test, minimising
\begin{equation}
    \chi^2 = \sum_i \frac{(n_{i,obs} - n_{i,model})^2}{n_{i,model}},
    \label{eq:chi_square_quadratic}
\end{equation}
where the dipole model is written as 
\begin{equation}
    n_{i,model} = \mathcal{M}(1 + \vec{d}\cdot\vec{\hat{n_i}}).
    \label{eq:dipole_model}
\end{equation}
Here the dipole amplitude on a given cell is given by the inner product between the dipole vector $\vec{d}$ and the unit vector pointing in the direction of the cell $\vec{\hat{n_i}}$, with $\vec{d}\cdot\vec{\hat{n_i}} = \mathcal{D}\cos\theta_i$. In addition to the dipole vector the monopole $\mathcal{M}$ is a free parameter, for which the mean value of all cells $\overline{n}$ is a good initial estimate. The $\chi^2$ test is agnostic to the actual distribution of the data, but a dipole model is imposed on the data. This can lead to misleading results if there are anisotropies in the data, such as those caused by systematic effects, on large enough scales and with large enough amplitude to influence the fit. This can generally be assessed with the reduced $\chi^2$, which should take a value around unity if the fit is good.

\subsection{Poisson estimator}
\label{sec:basic_poisson}

In the dipole estimators used in previous works there has been no explicit assumption on the shape of the distribution of sources. A Gaussian distribution can be a valid assumption for a source distribution, it has an additional degree of freedom compared to Poisson and is less valid if cell counts are low. However, as we do not know a priori how many sources we have nor commit to a cellsize for which we count sources, we choose to assume a Poisson distribution for our cell counts. The Poisson probability density function is given by Equation~\ref{eq:poisson}, which depends on the mean of the distribution $\lambda$, analogous to the monopole $\mathcal{M}$ in the absence of any anisotropies. To account for the effect of the dipole, we introduce a dipole model equivalent to Equation~(\ref{eq:dipole_model}), 
\begin{equation}
    \lambda(\vec{d},\mathcal{M}) = \mathcal{M}(1+\vec{d}\cdot\vec{\hat{n}}).
\end{equation}
In order to estimate the dipole parameters, we maximise the likelihood, which is given by
\begin{equation}
    \mathcal{L}(n|\vec{d},\mathcal{M}) = \prod_i  \frac{\lambda(\vec{d},\mathcal{M})^{n_i} e^{-\lambda(\vec{d},\mathcal{M})}}{n_i!}.
    \label{eq:poisson_likelihood}
\end{equation}
Maximising the likelihood through posterior sampling has the key advantage of immediately yielding the uncertainties on the derived parameter values. This removes the necessity for null hypothesis simulations as performed in e.g. \citet{Rubart2013,Secrest2021,Secrest2022}. This estimator was recently used by \citet{Dam2022}, which was shown to provide more tight constraints on the dipole than previously used methods. 

\subsection{Poisson-RMS estimator}
\label{sec:rms_poisson}

While a survey can be influenced by many different systematic effects, we can assert that the net effect is different sensitivity of the survey at different parts of the sky, leading to anisotropic number counts. Therefore we assume that all systematics that impact source counts in fact influence local noise, thereby causing the source density to vary across the survey. If the survey has sensitivity information in each part of the sky, the impact can be simply modelled by introducing additional variables to the model. So long as a detection threshold is consistently applied to the entire survey, the lower flux density limit will be linearly related to the local RMS noise. Consequently, taking into account the dipole and the power law describing number counts, we can model the mean counts in the Poisson estimator as
\begin{equation}
    \lambda(\vec{d}, \mathcal{M}, \sigma, x) = \mathcal{M}\left(\frac{\sigma}{\sigma_0}\right)^{-x}(1+\vec{d}\cdot\vec{\hat{n}}),
    \label{eq:lambda_rms}
\end{equation}
where $\sigma$ is the RMS noise of the cell, $x$ is the power-law index of the flux distribution, and $\sigma_0$ is a reference RMS value which scales the power law and explicitly ensures $\lambda$ is dimensionless. The value of $\sigma_0$ does not influence any parameters except for the monopole $\mathcal{M}$, which will take a value closest to the mean cell count $\overline{n}$ when taking $\sigma_0$ as equal to the median RMS noise over all cells. As the dipole amplitude depends on the the power law index of the flux distribution near the flux limit, the variation in the RMS noise should be small enough that it can be adequately described with a single value. Expected is a linear relation between the flux density limit and the local noise, as the detection threshold for most surveys is some multiple of the noise, usually $5\sigma$. Maximising the likelihood given by Equation~\ref{eq:poisson_likelihood} while inserting Equation~\ref{eq:lambda_rms} for $\lambda$ can thus yield the best-fit dipole and power-law parameters.

\subsection{Multi-Poisson estimator}
\label{sec:multi_poisson}

Hoping to remove any systematic effects stemming from the incomplete sky coverages of individual radio surveys, \citet{Darling2022} combined the Rapid Australian Square Kilometre Array Pathfinder (ASKAP) Continuum Survey \citep[RACS,][]{McConnell2020} and the VLA Sky Survey \citep[VLASS,][]{Lacy2020}, surprisingly finding a dipole that agrees with CMB in both amplitude and direction, though with large uncertainties. \citet{Secrest2022} note two inherent problems to this approach. Not only might selecting catalogues at different frequencies select for different spectral indices invalidating the assumption of a common dipole amplitude, but combining the catalogues in such a way ignores systematics that can vary between catalogues. Indeed it is most likely the second factor that plays the most important role, as things like observing frequency, array configuration, and calibration can all impact the number counts within a survey in ways that are difficult to predict. This is even more true for independent surveys.

Keeping that in mind, we approach the combination of any two catalogues in a different way. We will not make any attempt to unify the catalogues by matching, smoothing or creating a common map. Rather, we take both catalogues as independent tracers of the same dipole, allowing both catalogues to have a different monopole amplitude $\mathcal{M}$. In the Poisson estimator we thus estimate $\mathcal{M}$ separately for both catalogues, turning the likelihood into
\begin{equation}
    \label{eq:poisson_likelihood_multi}
    \begin{split}
    \mathcal{L}(n_1,n_2|\vec{d},\mathcal{M}_1, \mathcal{M}_2) & = \prod_i \frac{\lambda(\vec{d},\mathcal{M}_1)^{n_{1,i}}e^{-\lambda(\vec{d},\mathcal{M}_1)}}{n_{1,i}!}  \\
    &\times \prod_j \frac{\lambda(\vec{d},\mathcal{M}_2)^{n_{2,j}}e^{-\lambda(\vec{d},\mathcal{M}_2)}}{n_{2,j}!}.  
    \end{split}
\end{equation}
This likelihood benefits if the (expected) dipole amplitudes of both catalogues are similar. Any differences however will be absorbed into the overall error budget by virtue of the sampling algorithm. Once again, the likelihood can be maximised through posterior sampling, yielding the best-fit dipole results as well as the monopoles $\mathcal{M}_1$ and $\mathcal{M}_2$ for both surveys.

\subsection{Priors and injection values}

For efficient parameter estimation through posterior sampling, proper priors must be set. While priors on parameters can take on many shapes based on prior knowledge, we take flat priors on all parameters. This only leaves us to define the extents of the probed parameter space, as well as the initial guesses to serve as a starting point for the posterior sampling. In terms of dipole parameters, we separately infer dipole amplitude $\mathcal{D}$, as well as the right ascension and declination of the dipole direction. We expect the dipole amplitude to take values of around $10^{-2}$, but to allow for more variation the prior on the dipole amplitude is $\pi(\mathcal{D}) = \mathcal{U}(0,1)$. Any point in the sky can represent the dipole direction, so logically the priors cover the entire sky; $\pi(\mathrm{R.A.}) = \mathcal{U}(0,360)$ and $\pi(\mathrm{Dec.}) = \mathcal{U}(-90,90)$. As an initial guess for these parameters we inject the approximate expected dipole parameters from the CMB, $\mathcal{D}=4.5\times10^{-3}$, $\mathrm{R.A.} = 168\degree$, $\mathrm{Dec.} = -7\degree$.

Additionally the parameters of the distribution of number counts are also estimated. In the basic Poisson case this is represented by the monopole $\mathcal{M}$. As the dipole is not expected to meaningfully impact this value, a good initial guess of the monopole is the mean of all cell counts $\overline{n}$. As the real monopole is likely close to this value, we choose the prior $\pi(\mathcal{M}) = \mathcal{U}(0,2\overline{n})$. For the Poisson-RMS estimator, both a monopole $\mathcal{M}$ and power-law index $x$ are estimated. Before the estimation we fit a power law to the cell counts to get initial estimates $\mathcal{M}_{init}$ and $x_{init}$, which also function as the initial guesses for these parameters. The initial monopole estimate informs the prior as we use $\pi(\mathcal{M}) = \mathcal{U}(0, 2\mathcal{M}_{init})$. For the power-law index $x$ a value around 0.75-1.0 is always expected, so we take the prior $\pi(x) = \mathcal{U}(0,3)$. 

\section{Data}
\label{sec:data}

Given the estimators introduced in Section~\ref{sec:estimators}, there are a multitude of available radio catalogues to possibly make use of for a dipole estimate. To keep the approach focused, we use NVSS and RACS, two catalogues that when combined cover the full sky, but have been processed in very different ways given the respective eras in which they have been produced. Given the introduction of novel estimators, the NVSS is the logical choice for verification, providing a baseline as the most thoroughly studied catalogue in terms of dipole measurements. The choice of RACS for the second catalogue is straightforward. Not only does it complement NVSS in terms of sky coverage, the inclusion of sensitivity information in the catalogue makes it suitable for testing the Poisson-RMS estimator described in Section~\ref{sec:rms_poisson}. The complementary sky coverage of the two catalogues also provides the best testing ground for the Multi-Poisson estimator described in Section~\ref{sec:multi_poisson}.

\begin{figure*}
    \centering
    \includegraphics[width=0.48\textwidth]{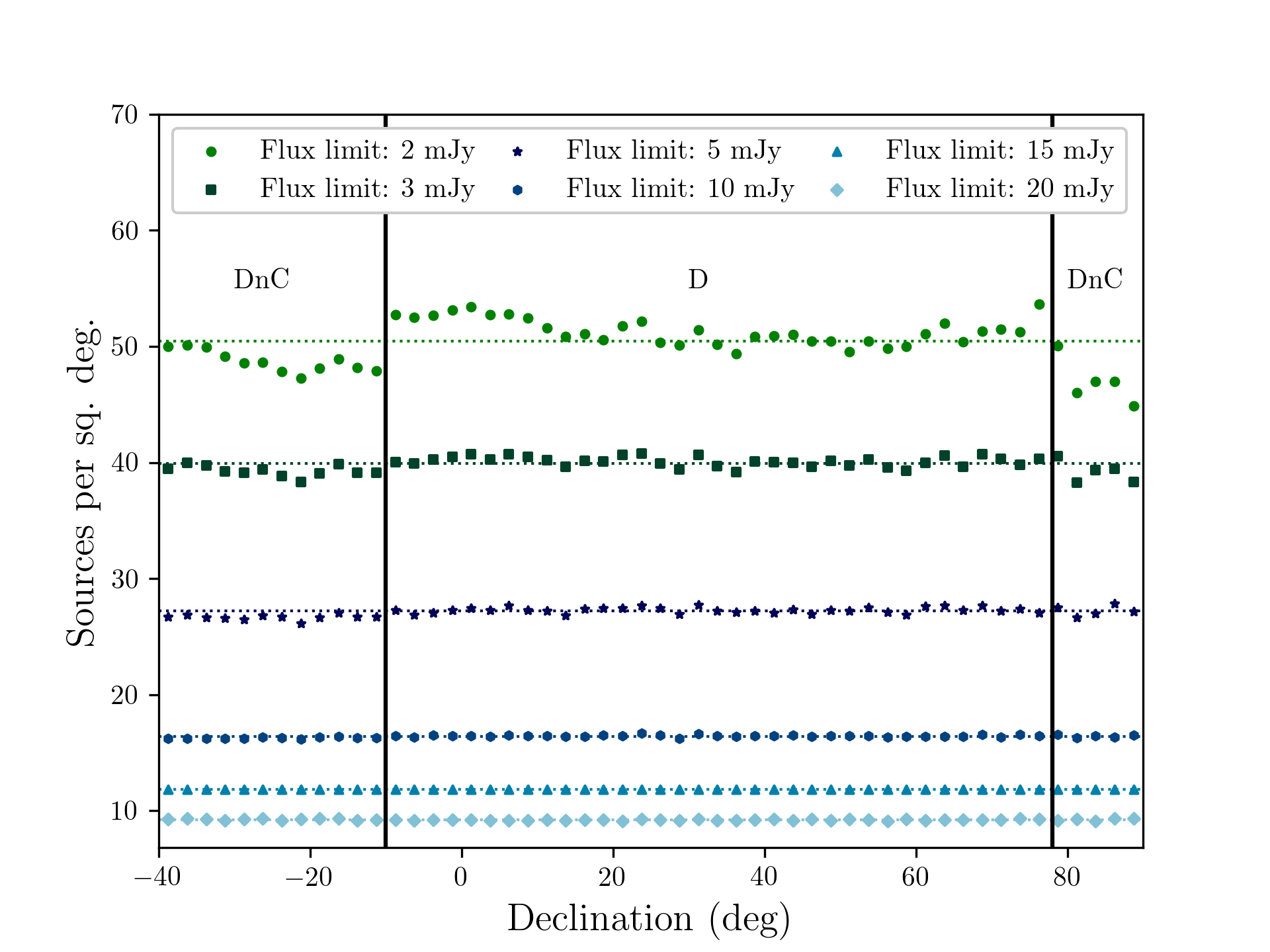}
    \includegraphics[width=0.48\textwidth]{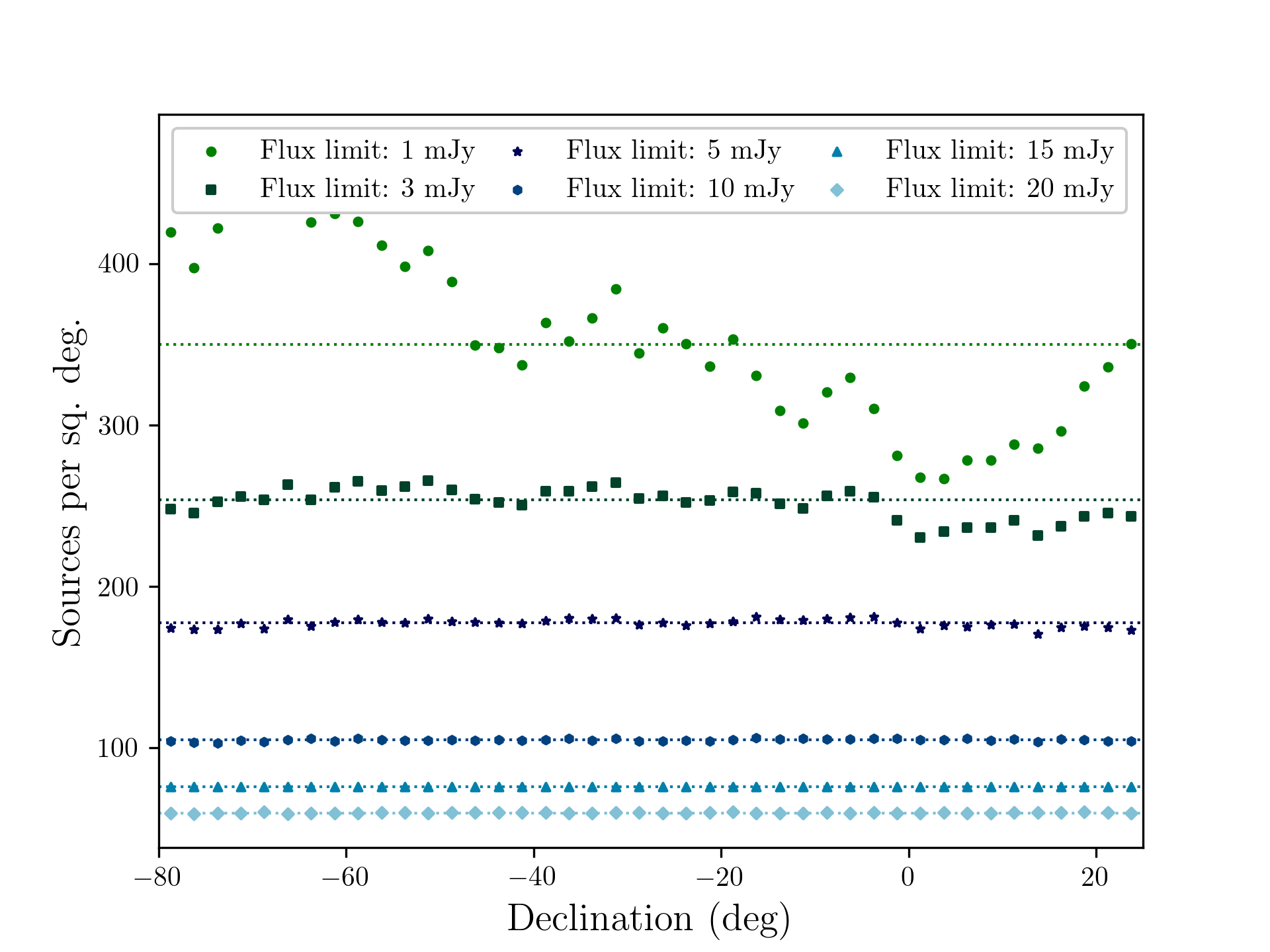}
    \caption{Source density of NVSS (left) and RACS (right) as a function of declination for different flux density cuts. For the NVSS is shown the boundaries where different array configurations are used. In both cases the catalogues are evidently inhomogenous at low flux densities.}
    \label{fig:dec_source_density}
\end{figure*}

\subsection{NVSS}

\begin{table}
    \centering
    \caption{Areas in NVSS masked due to high source density.}
    \begin{tabular}{c | c c c c c}
    Region & $\mathrm{RA_{min}}$ & $\mathrm{RA_{max}}$ & $\mathrm{Dec_{min}}$ & $\mathrm{Dec_{max}}$ & Sky area \\
           & \degree  & \degree  & \degree  & \degree & sq. deg. \\
    \hline \hline 
    1  & 82.0      & 90.0     & -7.0     & -1.0    & 47.9 \\
    2  & 49.0      & 52.0     & -39.0    & -36.0   & 7.1 \\ 
    3  & 185.0     & 189.0    & 11.0     & 14.0    & 11.7 \\
    \hline
    Total &           &          &         &          & 66.7 \\
    \end{tabular}
    \label{tab:nvss_mask}
\end{table}

The NVSS \citep{Condon1998} is one of the most well-studied surveys when it comes to dipole measurements, and as such is well suited for verifying novel dipole estimators. It covers the whole sky north of $-40\degree$ declination, has a central frequency of 1.4 GHz, with an angular resolution of 45\arcsec. The complete catalogue includes the Galactic plane and contains 1,773,484 sources.

An important feature of the NVSS catalogue is that for observations below a declination of $-10\degree$ and above a declination $78\degree$, the VLA DnC array configuration was used for observations, where for the rest of the survey VLA D configuration was used. This affects the number counts at those declinations. The left plot of Figure~\ref{fig:dec_source_density} shows the source density of NVSS as a function of declination for different flux density cuts. The impact of the different array configurations can be clearly seen at lower flux densities. Only around 15 mJy the source density becomes homogeneous, so for an unbiased dipole analysis we choose to exclude all sources with a flux density below 15 mJy, which is a commonly applied flux density cut \citep[e.g.][]{Singal2011,Siewert2021}. Even after this cut, some areas with significantly high source counts are present in the data. We mask these areas as specified in Table~\ref{tab:nvss_mask} for the dipole estimate.

With a flux density cut at 15 mJy, we fit a power law to the lower end of the flux distribution of sources and find a power-law index of $x = 0.85$. Additionally taking $\alpha=0.75$ and taking the velocity from the CMB into account (Equation~\ref{eq:dipole}), this sets the expectation of the dipole amplitude to $\mathcal{D} = 4.30\times10^{-3}$.

\subsection{RACS}

\begin{table}
    \centering
    \caption{Areas in RACS masked due to low source density.}
    \begin{tabular}{c | c c c c c}
    Region & $\mathrm{RA_{min}}$ & $\mathrm{RA_{max}}$ & $\mathrm{Dec_{min}}$ & $\mathrm{Dec_{max}}$ & Sky area \\
           & \degree  & \degree  & \degree  & \degree & sq. deg. \\
    \hline \hline 
    1     & 357.0     & 3.0      & 16.0     & 22.0    & 34.0 \\
    2     & 330.0     & 337.0    & 16.0     & 22.0    & 39.7 \\ 
    3     & 252.0     & 261.0    & 3.0      & 9.0     & 53.7 \\
    4     & 184.0     & 192.0    & 9.5      & 15.5    & 46.8 \\
    \hline
    Total &           &          &         &          & 174.2 \\
    \end{tabular}
    \label{tab:racs_mask}
\end{table}

RACS is the first large survey carried out using the Australian Square Kilometre Array Pathfinder (ASKAP), covering the sky south of $+40\degree$ declination. Observations are carried out with a central frequency of 887.5 MHz and images are smoothed to a common angular resolution of 25\arcsec. The first data release of RACS in Stokes I is described in \citet{Hale2021}, and the catalogue used in this work is the RACS catalogue with the Galactic plane removed, containing 2,123,638 sources. Source finding in the images has been done with the Python Blob Detector and Source Finder \citep[PyBDSF,][]{Mohan2015}, which provides a wealth of information on each source. Most importantly, the root-mean-square (RMS) noise at the position of each source is present in the \verb|noise| column of the catalogue, which we will utilise in the Poisson-RMS estimator approach.

The right plot of Figure~\ref{fig:dec_source_density} shows the source density of RACS as a function of declination for different flux density cuts. Even though there is no change in array configuration as with NVSS, there is a clear gradient in source density, where a lot more sources are detected at lower declinations. Once again, the catalogue becomes homogeneous at flux densities of around 15 mJy, so we exclude all sources below this flux density, as with NVSS. Even after this cut, some areas with significantly low source counts are present in the data. These areas appear just above the celestial equator. We mask these areas as specified in Table~\ref{tab:racs_mask} for the dipole estimate. 

With a flux density cut at 15 mJy and the other masks applied, we fit a power law to the lower end of the flux distribution of sources of RACS and find a power-law index of $x = 0.82$. Taking once again $\alpha=0.75$, this sets the expectation of the dipole amplitude to $\mathcal{D} = 4.24\times10^{-3}$.

\subsection{Common masks and pixelation}

\begin{figure*}
    \centering
    \includegraphics[width=0.48\textwidth]{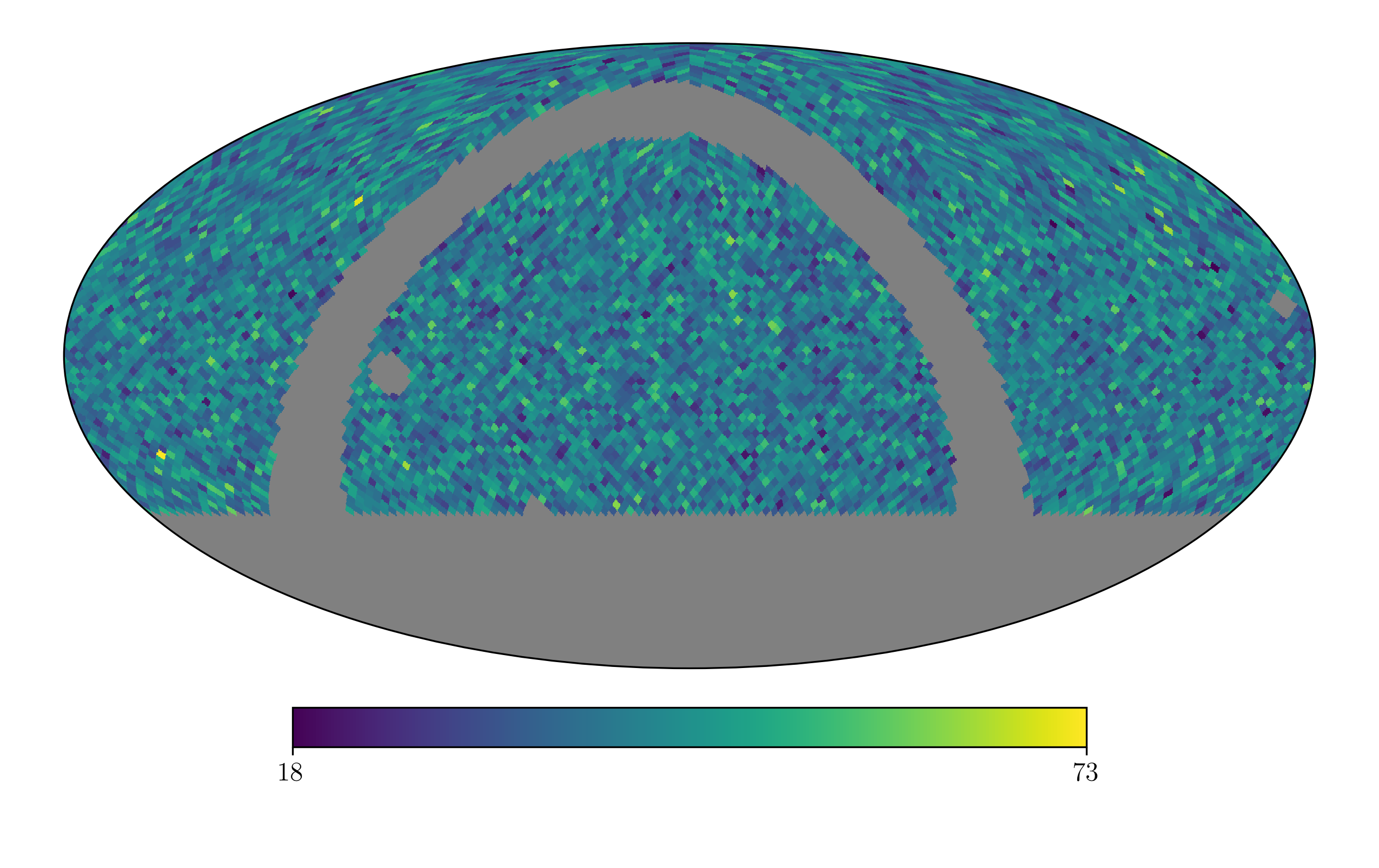}
    \includegraphics[width=0.48\textwidth]{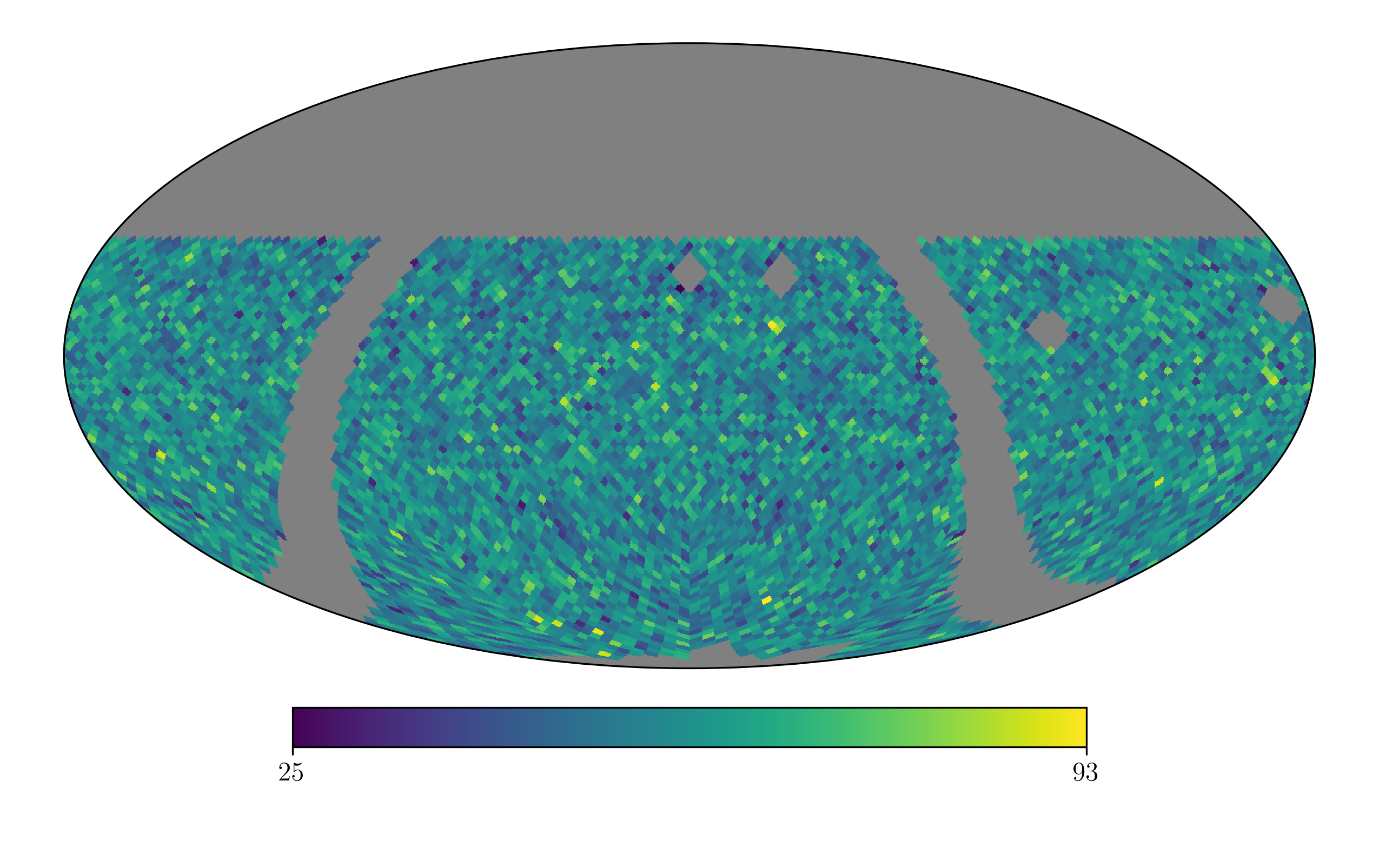}
    \caption{Number counts for NVSS (left) and RACS (right) in equatorial coordinates, in $N_{side} = 32$ \textsc{HEALPix} maps with masks and a flux density cut of 15 mJy applied.}
    \label{fig:nside32_healpix}
\end{figure*}

In order to avoid biases to the data, we adopt a masking scheme that uses the same principles for both NVSS and RACS. As mentioned previously, surveys are generally not homogeneous at the lowest flux densities due to variations in the noise, so we must choose a lower flux density threshold appropriate for the survey. As described before, for both NVSS and RACS we choose a flux density threshold of 15 mJy. To avoid counting Galactic sources or Galactic extended emission, we exclude the Galactic plane. Due to problems with source finding in the Galactic plane, it is already removed in the RACS catalogue, excluding all sources with $|b| < 5\degree$. For NVSS, overdensities from the galactic plane extend further out, and we exclude all sources with $|b| < 7\degree$. Finally we expect that the brightest sources in a survey push to the limit of the dynamic range, which introduces artefacts around bright sources. Due to differences in source finding methods between the two catalogues, this increases counts around bright sources in NVSS, but decreases counts around bright sources in RACS. In both cases the local source density is affected, and as such we remove all sources within a radius of $0.3\degree$ distance to any source brighter than 2.5 Jy. 

After masking data, we use the Hierarchical Equal Area isoLatitude Pixelation \citep[\textsc{HEALPix},][]{Gorski2005}\footnote{\url{http://healpix.sourceforge.net}} scheme to divide the sky into equal sized cells. \textsc{HEALPix} allows the flexibility of choosing the number of cells over the whole sky, with the base and minimum value being 12 pixels. The resolution parameter $N_{side}$ determines the number of pixels by $N_{pix} = 12 \times N_{side}^2$. We choose two resolutions for our experiment, $N_{side}=32$ and $N_{side}=64$, which have pixel sizes of 110\arcmin\ and 55\arcmin on a side, respectively. For measuring number counts, each cell holds the number of sources detected within the confines of the cell. To avoid edge effects resulting from pixels covering data only partially, all pixels that have a neighbouring pixel with zero sources are also set to zero. Finally, all pixels with a value of zero are masked to ensure that these pixels are not taken into account during dipole estimation. The \textsc{HEALPix} maps of NVSS and RACS with $N_{side}=32$, with flux density cuts and masks applied are shown in Figure~\ref{fig:nside32_healpix}. For the Poisson-RMS estimator, the local noise is determined per \textsc{HEALPix} cell by taking the median RMS of all the sources within the cell.

\subsection{Simulations}

In addition to the survey data sets of NVSS and RACS, we create a catalogue of simulated sources with a dipole effect to test of the validity of the estimators. To do this, we uniformly populate the sky with sources, and assign a rest flux density $S_{rest}$ according to a power law
\begin{equation}
    S_{rest} = S_{low}(1 - \mathcal{U})^{-1/x},
\end{equation}
where $\mathcal{U}$ is a uniform distribution between 0 and 1. $S_{low}$ is the lower flux density limit at which sources are generated, and $x$ the power law index of the flux density distribution. We transform the rest flux densities and positions of sources by applying relativistic aberration, Doppler shift and Doppler boost as expressed in Equations~\ref{eq:dipole_flux} and \ref{eq:dipole_pos}. 

We add Gaussian noise to the flux densities of the sources, generating a larger sample of sources and simulating source extraction by only including sources with S/N > 5 in the final catalogue. This naturally adds the effect of Eddington bias to the sample, which is expected to be present in real source catalogues. For a realistic distribution the local noise variation is taken from the RACS $N_{side}=32$ RMS map, idealising by assigning the RMS of a cell to all sources in that cell. Additionally, we simulate false detections by generating sources with the same flux distribution that are not affected by the dipole. These sources consist of 0.3\% of the total catalogue, which is the percentage reported for the RACS catalogues \citep{Hale2021}.

All sources are simulated with a spectral index of 0.75, and the power law used to generate the flux distribution has $x=1$. We apply the dipole effect assuming the direction derived from the CMB dipole, (RA,Dec.)$ = (170\degree,-10\degree)$, but with an increased velocity of $v=1107$ km/s, to ensure that similar sensitivity to the dipole as NVSS and RACS will be reached,  with similar monopole values. This sets the expectation of the dipole to $\mathcal{D} = 1.5\times10^{-2}$. 

\section{Results}
\label{sec:results}

\begin{table*}[]
    \renewcommand*{\arraystretch}{1.4}
    \centering
    \caption{Dipole estimates using the various estimators on the NVSS and RACS catalogues, including a combined estimate using both catalogues.}
    \resizebox{\textwidth}{!}{%
    \begin{tabular}{c | c | c c c | c c | c c c | c c}
    Survey & Estimator & NSIDE & $S_0$ & $N$ & $\mathcal{M}$ & $x$ & $\mathcal{D}$ & R.A. & Dec. & $\chi^2/dof$ \\
    &  & & (mJy) & & counts/pixel & &($\times10^{-2}$) & (deg)  & (deg) & \\
    \hline \hline
    Simulation & Poisson & 32 & 1 & 4,205,065 & $489.6\pm0.3$ & -- & $31.7\pm0.1$ & $202\pm1$ & $-78.5\pm0.2$ & 37.3\\
    &  & 32 & 15 & 434,247 & $54.5\pm0.1$ & -- & $1.50\pm0.25$ & $172\pm10$ & $-17\pm13$ & 1.01 \\
    &  & 32 & 50 & 130,150 & $16.3\pm0.1$ & -- & $2.06\pm0.60$ & $178\pm19$ & $-46_{-14}^{+17}$ & 1.02 \\
    & Poisson-RMS & 32 & $5\sigma$ & 4,212,472 & $518.9\pm0.3$ & $1.000\pm0.002$ & $1.58\pm0.08$ & $169\pm4$ & $-12\pm4$ & 1.10 \\
    \hline
    RACS & Quadratic & 32 & 15 & 451,003 & $57.4\pm0.1$ & -- & $1.35\pm0.26$ & $192\pm12$ & $6\pm17$ & 1.23 \\
    & & 64 & 15 & 458,152 & $14.74\pm0.03$ & -- & $1.30\pm0.24$ & $195\pm11$ & $4\pm16$ & 1.11 \\
    & Poisson & 32 & 15 & 451,003 & $56.8\pm0.1$ & -- & $1.41\pm0.24$ & $193^{+11}_{-10}$ & $4^{+14}_{-15}$ & 1.24 \\
    &  & 64 & 15 & 458,152 & $14.19\pm0.02$ & --  & $1.42\pm0.24$ & $194\pm10$ & $6\pm14$ & 1.13 \\
     & Poisson-RMS & 32 & 5$\sigma$ & 2,035,375 & $253.3\pm0.2$ & $0.778\pm0.003$ & $1.62\pm0.12$ & $210\pm4$ & $-12^{+7}_{-6}$ & 2.03 \\
    & & 64 & 5$\sigma$ & 2,068,204 & $63.8\pm0.1$ & $0.738\pm0.003$ & $2.0\pm0.2$ & $204\pm5$ & $-40\pm4$ & 1.49 \\
    \hline
    NVSS & Quadratic & 32 & 15 & 345,803 & $41.0\pm0.1$ & -- & $1.35\pm0.30$ & $145\pm13$ & $-5\pm17$ & 1.17 \\
    & & 64 & 15 & 352,862 & $10.66\pm0.02$ & -- & $1.23\pm0.29$ & $150\pm13$ & $-10\pm17$ & 1.09 \\
    & Poisson & 32 & 15 & 345,803 & $40.4\pm0.1$ & -- & $1.40\pm0.29$ & $146\pm12$ & $-5\pm15$ & 1.26 \\
    &  & 64 & 15 & 352,862 & $10.11\pm0.02$ & -- & $1.38\pm0.29$ & $151\pm12$ & $-10\pm15$ & 1.15 \\
    \hline
    & & & & & $\mathcal{M}_1$ & $\mathcal{M}_2$ & & & \\
    & & & & & counts/pixel & counts/pixel & & & \\
    \hline
    NVSS + RACS & Multi-Poisson & 32 & 15 & 796,806 & $40.4\pm0.1$ & $56.7\pm0.1$ & $1.30\pm0.18$ & $173\pm9$ & $-1\pm11$ & -- \\
    &   & 64 & 15 & 811,014 & $10.11\pm0.02$ & $14.18\pm0.02$ & $1.29\pm0.18$ & $175\pm8$ & $-2\pm11$ & -- \\
    \end{tabular}}
    \label{tab:dipole_results}
\end{table*}

\begin{figure}
    \centering
    \includegraphics[width=\hsize]{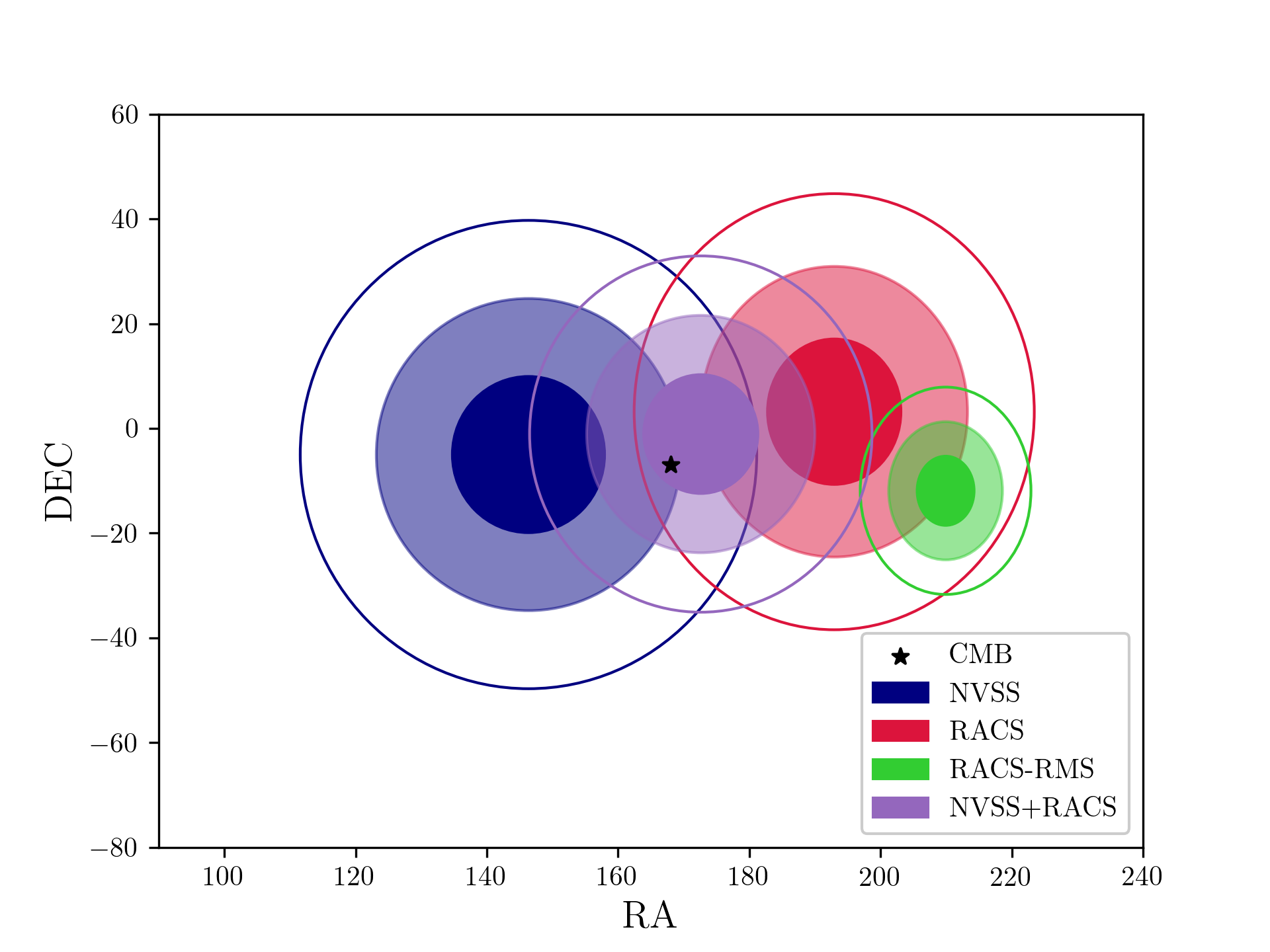}
    \caption{Best-fit dipole directions for the Poisson estimator of NVSS (blue), RACS (red), RACS with RMS power law (green), and NVSS+RACS (purple), compared with the CMB dipole direction (black star). Different transparency levels represent 1-,2- and $3\sigma$ uncertainties. In all cases results from the $N_{side}=32$ HEALPix map are shown.}
    \label{fig:dipole_dirs}
\end{figure}

\begin{figure}
    \centering
    \includegraphics[width=\hsize]{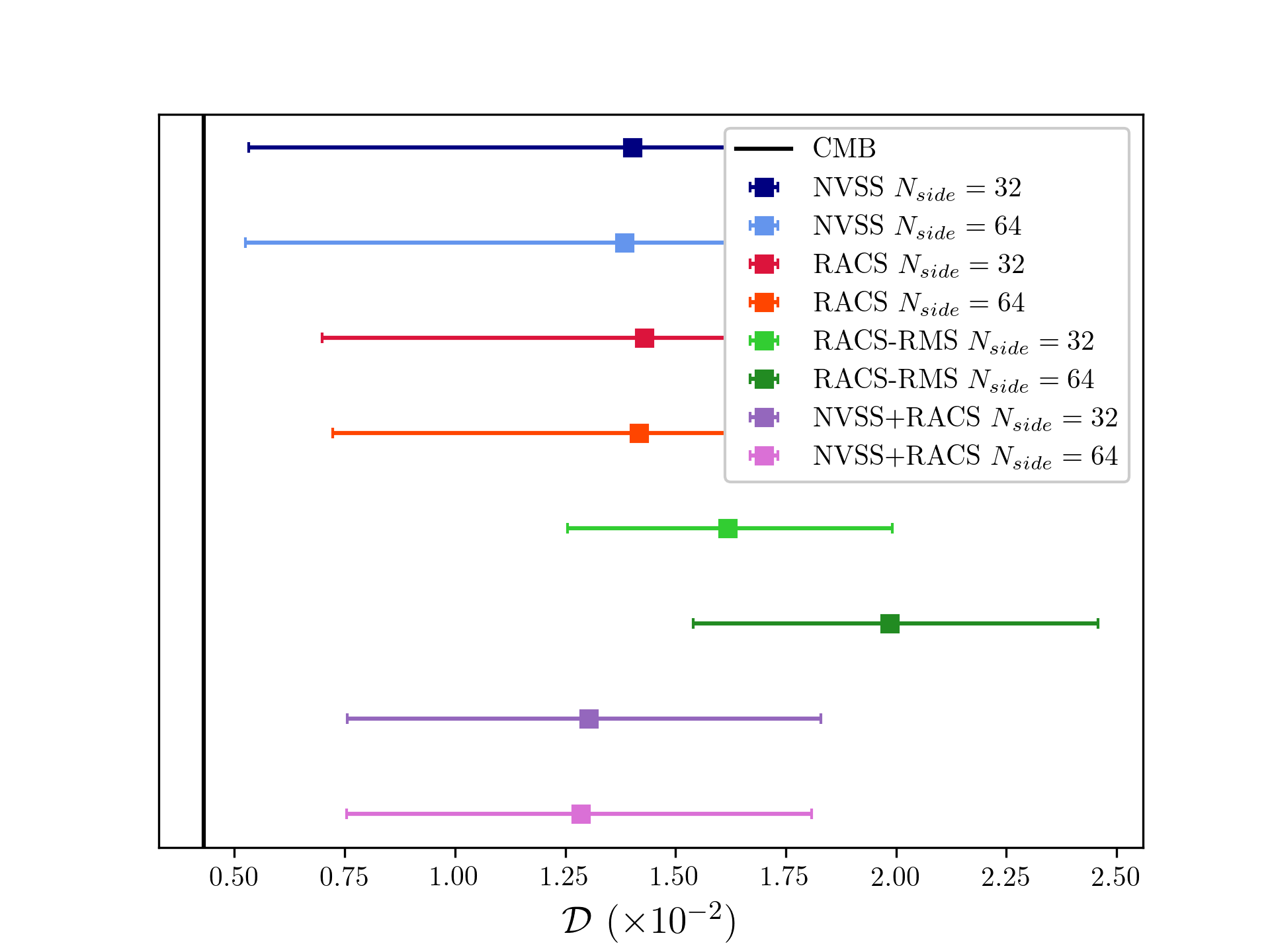}
    \caption{Best-fit dipole amplitudes with $3\sigma$ uncertainties for the Poisson estimator of NVSS (blue), RACS (red), RACS with RMS power law (green), and NVSS+RACS (purple), compared with an expected CMB dipole amplitude of $\mathcal{D} = 4.5\times10^{-3}$ (black line). Both results from the $N_{side}=32$ and $N_{side}=64$ \textsc{HEALPix} maps are shown.}
    \label{fig:dipole_amps}
\end{figure}

Using the described estimators, we estimate the dipole parameters for NVSS and RACS. The results are summarised in Table~\ref{tab:dipole_results}, with dipole directions shown in Figure~\ref{fig:dipole_dirs} and dipole amplitudes shown in Figure~\ref{fig:dipole_amps}. To estimate the best-fit parameters, for the quadratic estimator we minimise $\chi^2$ using \textsc{lmfit} \citep{Newville2016}. The reduced $\chi^2$-values are reported in Table~\ref{tab:dipole_results}. For the Poisson estimators we use the Bayesian inference library \textsc{bilby} \citep{Ashton2019}, which provides a convenient and user-friendly environment for parameter estimation. We maximise the likelihood using Markov chain Monte Carlo (MCMC) sampling with \textsc{emcee} \citep{Foreman-Mackey2013}. The scripts used to obtain these results are available on GitHub\footnote{\url{https://github.com/JonahDW/Bayesian-dipole}} and an immutable copy is archived in Zenodo \citep{Wagenveld2023a}. 

To get an indication of how well the distribution fits a Poisson distribution, Pearson's $\chi^2$ can also be used as a Poisson dispersion statistic, defined as
\begin{equation}
    \chi^2 = \sum_i\frac{(n_i - \overline{n})^2}{\overline{n}}.
    \label{eq:chi_square_poisson}
\end{equation}
As we use the mean of the distribution instead of a model expectation, the resulting $\chi^2$ value indicates the ratio between the variance and the mean of the distribution\footnote{\citet{Peebles1980} uses this measure as a clustering statistic of the large scale structure, used to define the number of objects per cluster.}, and thus for a Poisson distribution the $\chi^2$ value divided by the number of degrees of freedom ($dof$), $\chi^2/dof$, should be (close to) unity. In the case of the Poisson-RMS estimator, number counts are corrected for the derived power law before calculating $\chi^2$.

\subsection{Quadratic and Poisson estimators}

Both the quadratic and Poisson estimators are insensitive to gaps in the data, but will be sensitive to inhomogeneous source counts in the data. As such, we perform the flux density cuts on all the data before estimating the dipole. As specified in Section~\ref{sec:data}, we choose a flux density cut of 15 mJy for all catalogues in addition to the other masks detailed. As the simulated catalogue uses the local noise information from RACS, the same masking is applied there. This leaves ${\sim}3.5\times10^5$, ${\sim}4.5\times10^5$, and ${\sim}2.2\times10^6$ sources for NVSS, RACS, and the simulated catalogue respectively. The resulting number counts for NVSS and RACS are shown in Figure~\ref{fig:nside32_healpix}. Along with estimating the dipole amplitude and direction, we estimate the monopole $\mathcal{M}$. For NVSS and RACS these parameters are estimated using \textsc{HEALPix} maps of both $N_{side}$ 32 and 64, the results of the different cell sizes are shown in Table~\ref{tab:dipole_results}. 

The best-fit parameters for the simulated data set are shown in Table~\ref{tab:dipole_results}, both for a low threshold of 1 mJy, the common threshold of 15 mJy, and a high threshold of 50 mJy. As the noise variation of the simulated catalogue is based on RACS, the 15 mJy threshold should be appropriate to get a good estimate of the injected dipole. The low threshold shows that the dominant anisotropy from the RACS noise, which dominates the dipole by three orders of magnitude, is mostly a declination effect, but a smaller effect in right ascension is also observed. With the 15 mJy threshold the injected values for the dipole are retrieved within the uncertainties. To see how the estimator reacts to a lack of sources, for the 50 mJy threshold the required number counts are not reached for a $3\sigma$ measurement of the dipole amplitude. This does not however introduce a bias, as values still match the injected values, albeit with large uncertainties.

Comparing results between the quadratic and basic Poisson estimators on NVSS and RACS, values match within the uncertainties for all estimated parameters. For NVSS, the results of the quadratic estimator and Poisson estimator match those of \citet{Siewert2021} for a flux density cut of 15 mJy in terms of dipole amplitude, but the direction is slightly offset ($\Delta\theta\sim20\degree$). This is caused by a difference in masking strategy, as is already seen in \citet{Siewert2021} that different masks yield different dipole parameters. The low $\chi^2$ values for the Poisson estimators indicate that a Poisson assumption is in line with the expected distribution of source counts. 

As the results indicate in Table~\ref{tab:dipole_results} for both quadratic and Poisson estimators, the results between $N_{side}=32$ and $N_{side}=64$ pixel sizes agree with each other within the uncertainties. Furthermore, the dipole amplitudes of NVSS and RACS also agree within the uncertainties with each other, with the dipole amplitude from RACS being slightly higher. The dipole directions between RACS and NVSS are somewhat misaligned ($\Delta\theta\sim50\degree$), though both align with the CMB dipole direction within $3\sigma$, at $\Delta\theta\sim20\degree$ and $\Delta\theta\sim30\degree$ for NVSS and RACS respectively (see also Figure~\ref{fig:dipole_dirs}). Figure~\ref{fig:dipole_amps} shows the amplitudes including uncertainties of the results of the Poisson estimator on NVSS and RACS. In all cases, the amplitude of the dipole is 3-3.5 times higher than the dipole amplitude expectation from the CMB. For NVSS, the result is at $3.4\sigma$ significance and for RACS at a significance of $4.1\sigma$.

\subsection{Poisson-RMS estimator}

\begin{figure*}
    \centering
    \includegraphics[width=0.48\textwidth]{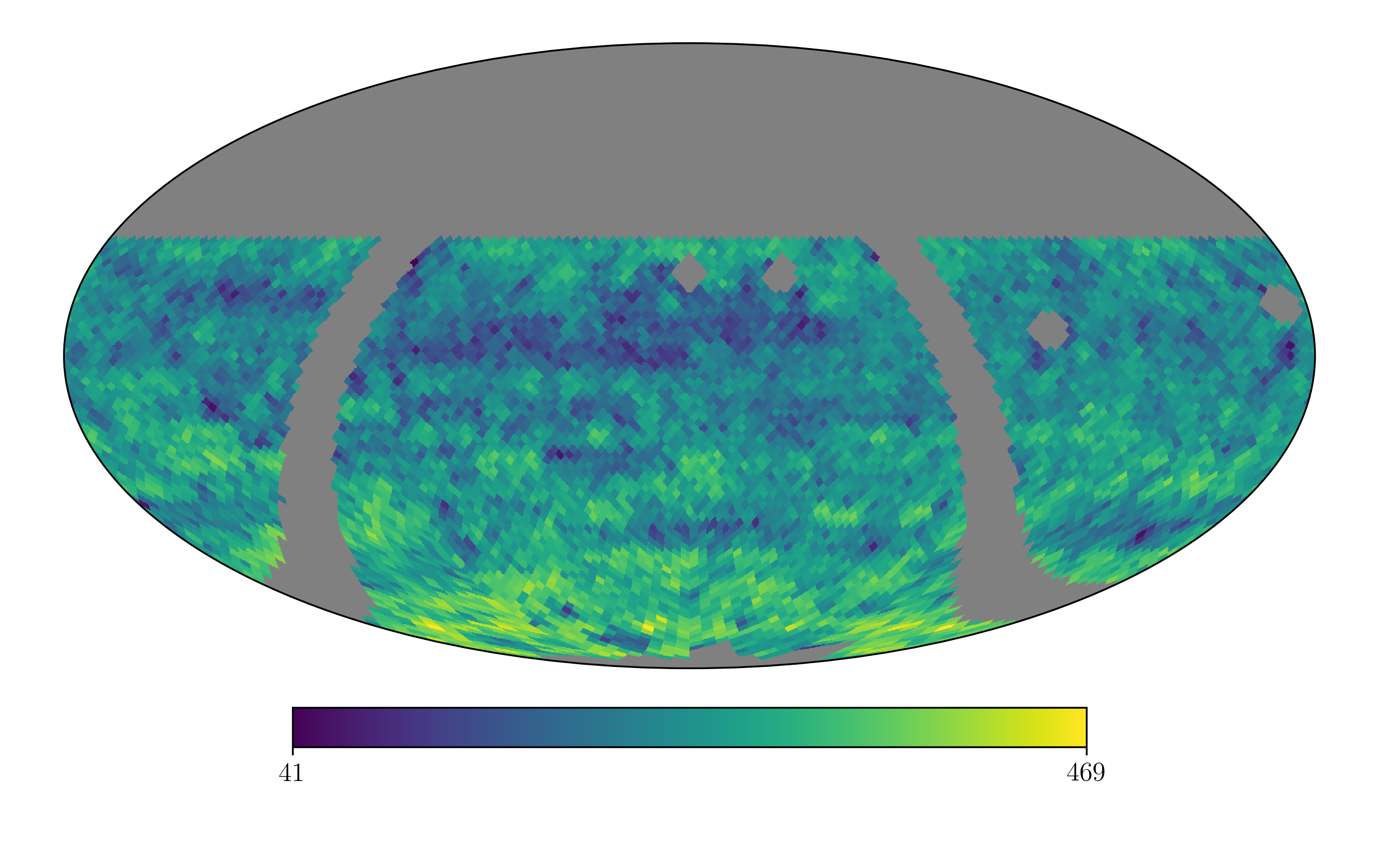}
    \includegraphics[width=0.48\textwidth]{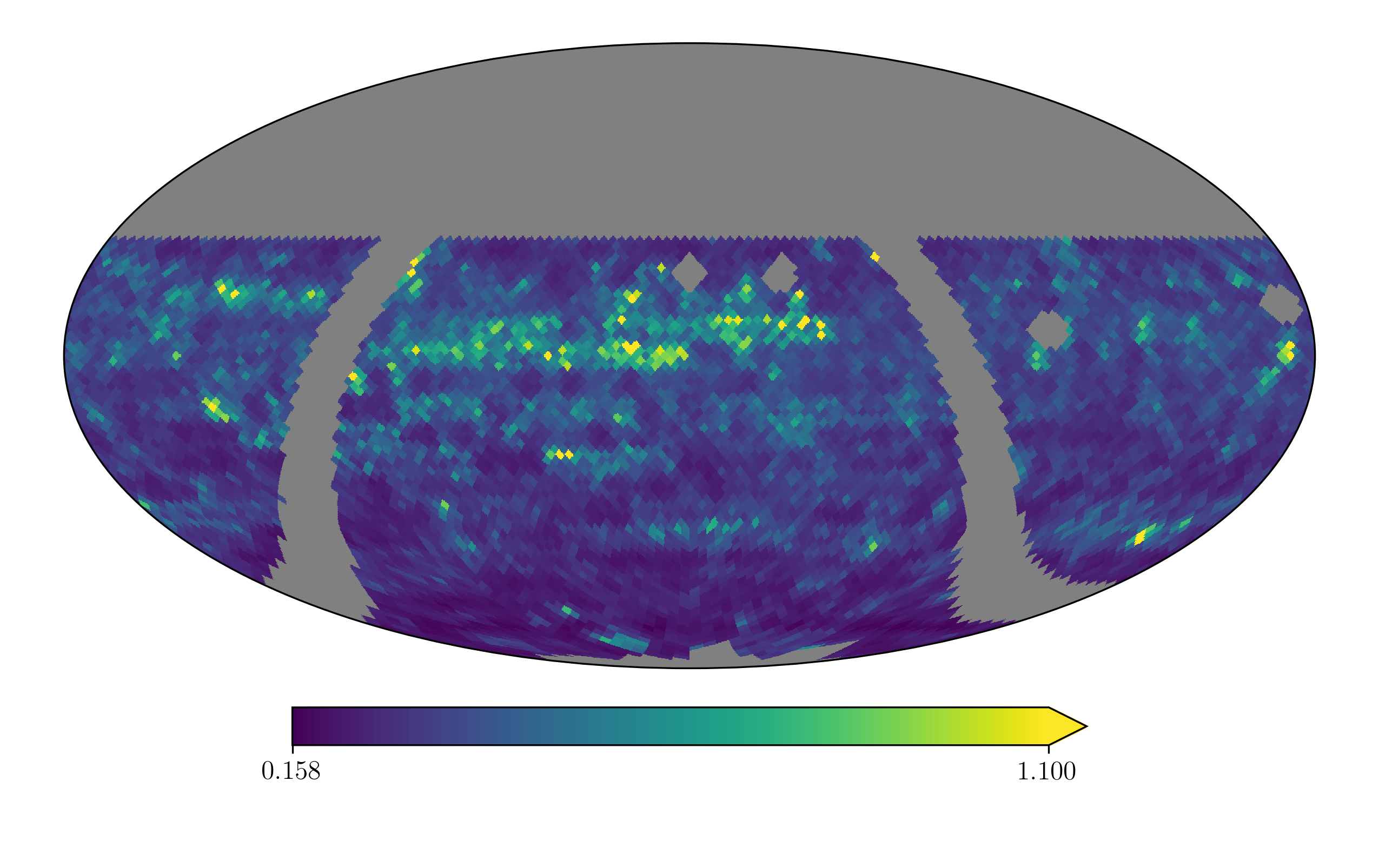}
    \caption{Number counts (left) and median RMS (right) for RACS in equatorial coordinates, in $N_{side} = 32$ \textsc{HEALPix} maps, with no flux density cuts applied.}
    \label{fig:nside32_racs}
\end{figure*}

\begin{figure*}
    \centering
    \includegraphics[width=0.48\textwidth]{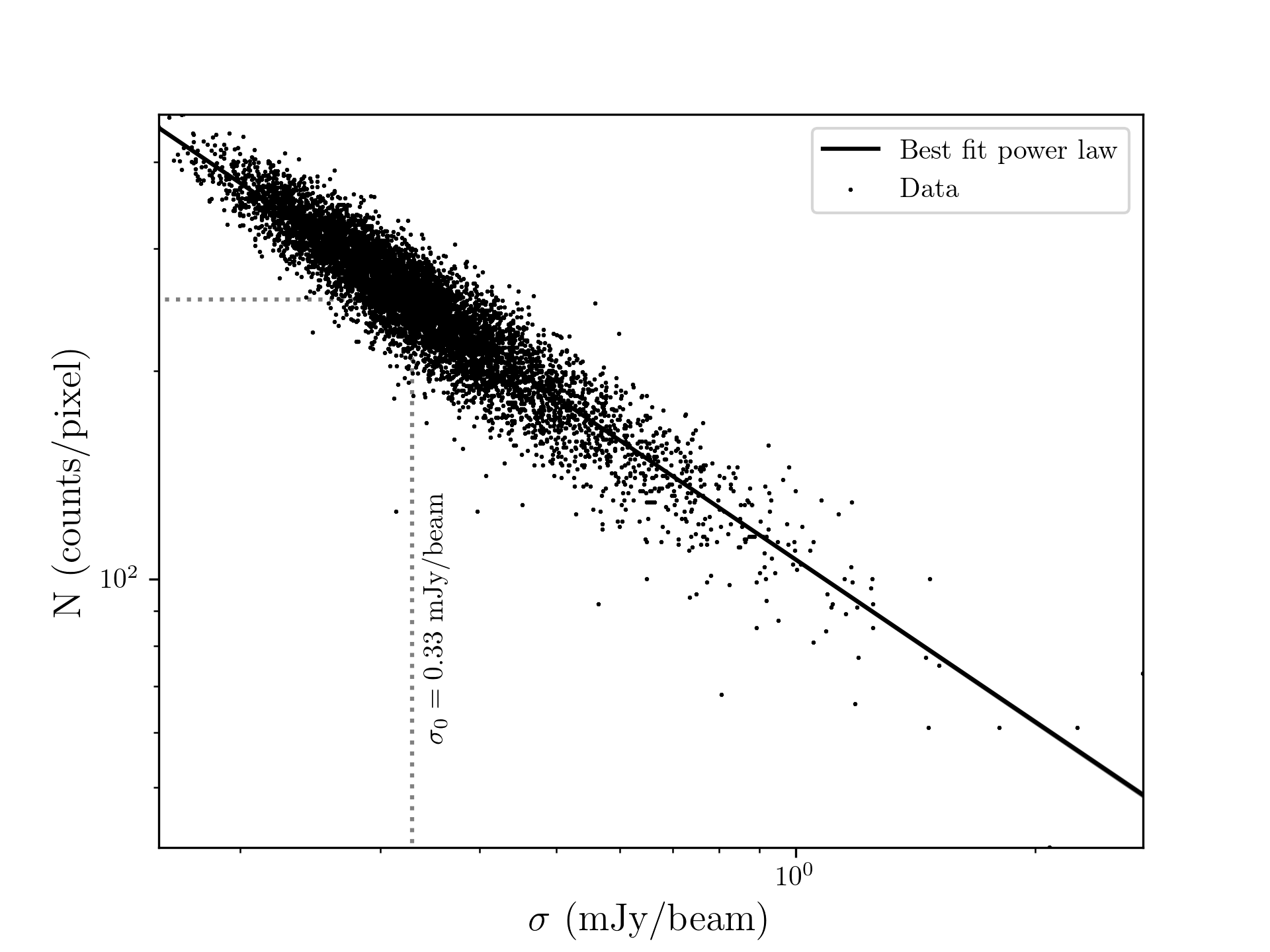}
    \includegraphics[width=0.48\textwidth]{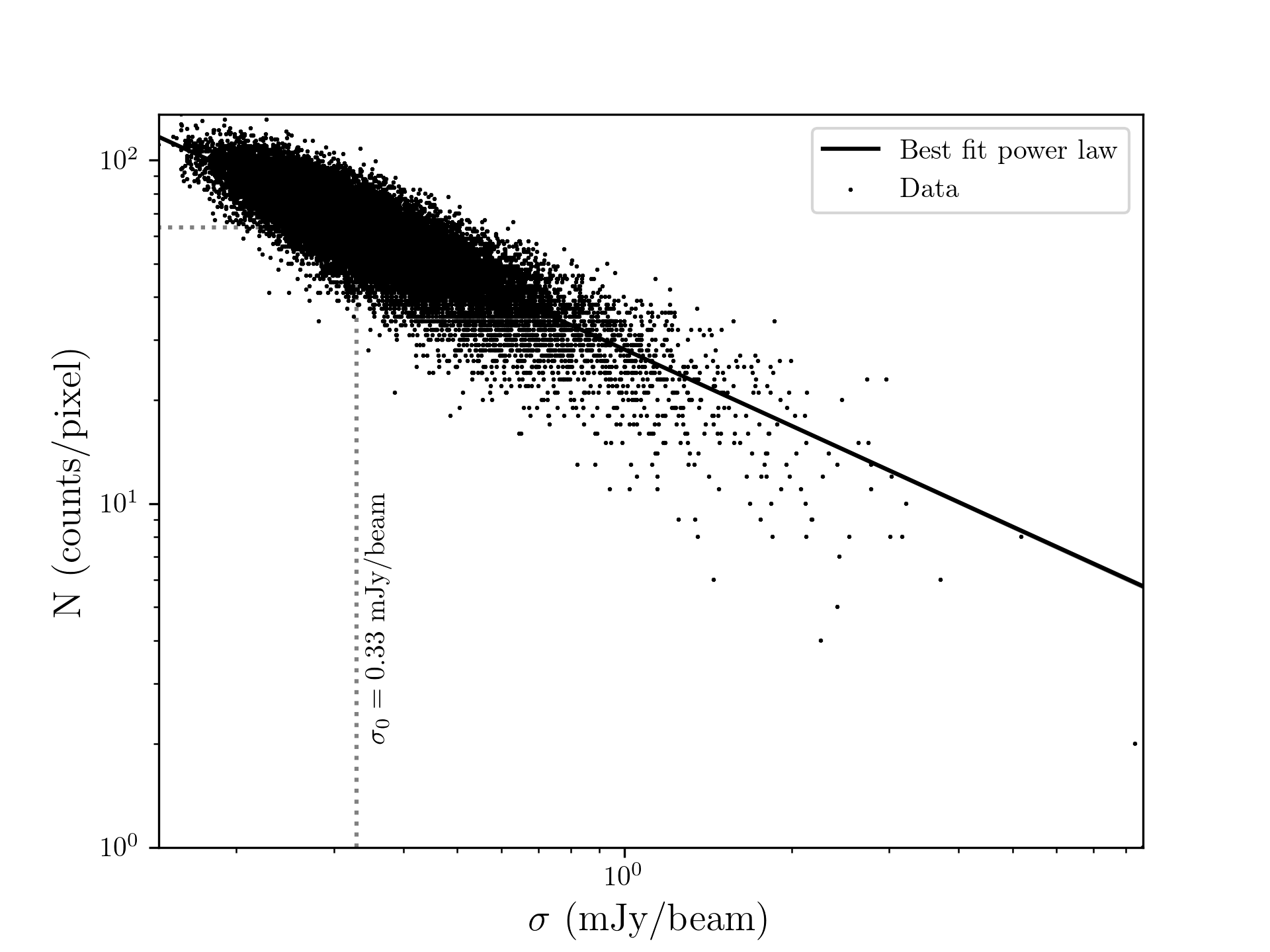}
    \caption{Cell counts of the $N_{side} = 32$ (left) and $N_{side} = 64$ (right) \textsc{HEALPix} maps with no flux density cuts applied, as a function of the median RMS of the pixels, along with the best fit power law model (black solid line). The determined $\sigma_0$ is indicated by the dashed vertical line, it intercepts the power law at the best fit monopole value, indicated by the horizontal dashed line.}
    \label{fig:powerlaw_fit}
\end{figure*}

As described in Section~\ref{sec:rms_poisson}, we aim to account for the variation in source counts across the survey by assuming these are described by the RMS noise of the images. For this estimator we don't apply the flux density cut, and instead, fit a power law that relates the RMS of a cell to the number counts in that cell. The RMS of the survey is not available for NVSS, but is present in the RACS catalogue for each source individually. We obtain the RMS of a cell by taking the median RMS value of all sources within it. We take the median RMS of all cells as the reference RMS, which is $\sigma_0=0.33$ mJy/beam. The \textsc{HEALPix} maps of source counts and median RMS per cell for RACS are shown in Figure~\ref{fig:nside32_racs}, showing the variation of RMS and source counts across the survey. Along with estimating the dipole parameters, the monopole $\mathcal{M}$ and power law index $x$ are estimated as well. For RACS, these parameters are estimated using \textsc{HEALPix} maps of both $N_{side}=32$ and $N_{side}=64$. 

The parameters for the simulated data set are estimated and shown in Table~\ref{tab:dipole_results}. The noise variation of the simulated catalogue is based on the $N_{side}=32$ RACS RMS map shown in Figure~\ref{fig:nside32_racs}, which in this case means that the RMS map is a perfect representation of the noise in the catalogue. The RMS estimator retrieves the injected dipole parameters, with a much higher significance than the standard Poisson estimator.

The results for RACS are shown in Table~\ref{tab:dipole_results}, showing a rather big discrepancy between the two pixel scales, and to other results as well. In both cases, the dipole amplitude is increased w.r.t. the quadratic and basic Poisson estimators, and the direction is no longer agreeing with the direction of the CMB dipole. The $N_{side}=32$ map seems to be less affected than the $N_{side}=64$ map, but in both cases the dipole direction is further away from the CMB dipole direction, with $\Delta\theta\sim40\degree$ and $\Delta\theta\sim45\degree$ separation for the $N_{side}=32$ and $N_{side}=64$ maps respectively. Especially striking is the recovered dipole direction of the $N_{side}=64$ map, which is at a declination of $-40\deg$. This retrieved dipole direction aligns towards the anisotropy retrieved in the simulated data with the 1 mJy flux density threshold. Rather than pointing to an additional systematic effect that is not modeled by the local RMS, it is therefore more likely that the median RMS noise per cell does not adequately represent the noise variation observed in the catalogue. 

To further investigate these results, the power law fits to the cells are shown in Figure~\ref{fig:powerlaw_fit}, indicating that both power laws are a good fit to the distribution. For the $N_{side}=64$ map the relation fits less well to the lower number count cells, possibly indicating that the power law assumption breaks down for cells with low number counts. One effect that can contribute to this is that at such low number counts the median RMS will be a less robust measure of the local noise. As with the other RACS results, there is a misalignment in right ascension that is even more pronounced here (see also Figure~\ref{fig:dipole_dirs}). As seen in Figure~\ref{fig:dipole_amps}, the dipole amplitude is also increased. For the $N_{side}=32$ map the dipole amplitude is 3.8 times higher than the CMB expectation with a formal significance of $10\sigma$, and for the $N_{side}=64$ map the dipole amplitude is 4.7 times higher with a formal significance of $8\sigma$. 

\subsection{Combining RACS and NVSS}
\label{sec:combine}

Following the procedure laid out in Section~\ref{sec:multi_poisson}, we do a combined estimate of the dipole parameters of NVSS and RACS, assuming a common dipole amplitude but independent monopole amplitudes. Although we have shown that slightly different dipole amplitudes are to be expected between the catalogues, the degree of this difference depends entirely on the degree to which the inferred dipole is kinetic. Regardless of this, due to the nature of parameter estimation with MCMC, any differences in dipole amplitude between the catalogues will be absorbed into the overall uncertainty of the estimated parameters. In terms of monopole, there is no question as to the difference between the catalogues, as from the estimates of the individual catalogues, even with the same cut in flux density, there are large differences in source density. 

Table~\ref{tab:dipole_results} shows the results of the dipole parameters for the combined estimate of NVSS + RACS. Whereas the dipole directions for the individual catalogues were misaligned with the CMB dipole direction, the combined estimate favours a dipole direction that is perfectly aligned ($\Delta\theta=4\degree$, see Figure~\ref{fig:dipole_dirs}) with that of the CMB dipole. Along with this, the dipole amplitude is reduced w.r.t. either of the individual catalogues, however it is still in tension with the CMB dipole. The dipole amplitude is three times higher than the CMB expectation, with a significance of $4.8\sigma$ for both the $N_{side}=32$ and $N_{side}=64$ maps. If we base our belief in a dipole result on its agreement with the CMB dipole in terms of direction, then this is the most significant and reliable result we obtain in this work. It is furthermore the most significant result obtained with radio sources to date, matching the significance of the dipole estimate with WISE AGN from \citet{Secrest2022}, although less significant than the joint WISE+NVSS result from the same work.

\section{Discussion}
\label{sec:discussion}

\begin{figure}
    \centering
    \includegraphics[width=\hsize]{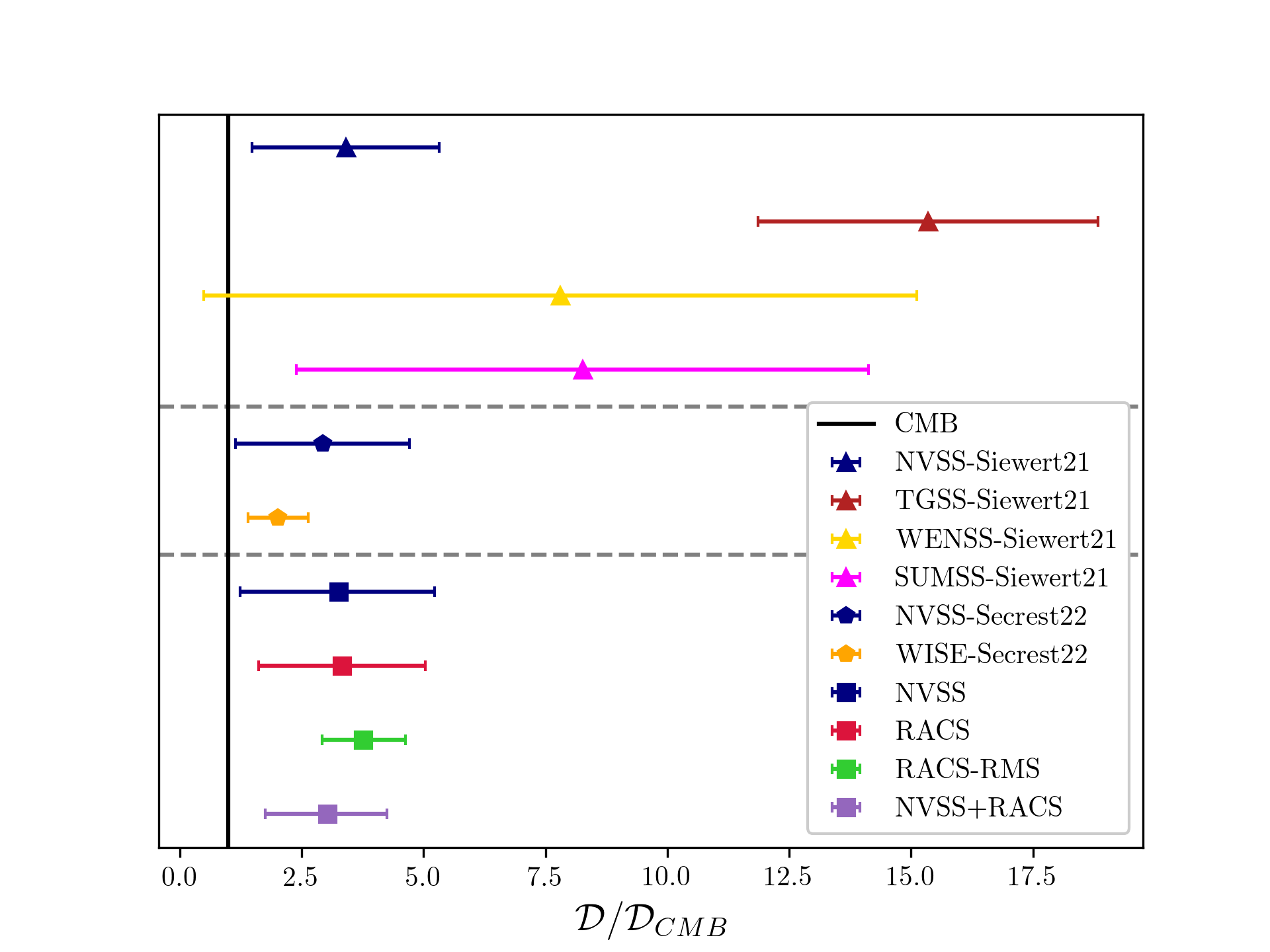}
    \caption{Dipole amplitudes with $3\sigma$ uncertainties compared to the amplitude expected from the CMB from this work and to results from \citet{Siewert2021} and \citet{Secrest2022}. The results from the different works are separated by horizontal dashed lines, showing the results from this work at the bottom.}
    \label{fig:dipole_amp_lit}
\end{figure}

The results presented here show the potential of our introduced estimators and present at the very least an alternative method of making dipole measurements in present and future surveys. Figure~\ref{fig:dipole_amp_lit} shows the results from this work compared to the most recent results from various surveys taken from \citet{Siewert2021} and \citet{Secrest2022} in terms of dipole amplitude. Between all works the NVSS is consistent, so too with our results. The results obtained here agree within uncertainties with both results from \citet{Secrest2022}, with the exception of RACS-RMS and the WISE result. The same goes for the results from \citet{Siewert2021}, with the exception of the TGSS result. Though the extremely high amplitude of TGSS is attributed to a frequency dependence of the dipole in \citet{Siewert2021}, the WISE result from \citet{Secrest2022} does not follow the fitted trend. \citet{Secrest2022} thus suggest that the TGSS result might deviate due to issues in flux calibration. As we have results at different frequencies we can tentatively check whether results match up with the frequency evolution model of the dipole amplitude from \citet{Siewert2021}, which predicts $\mathcal{D}=2.3\times10^{-2}$ at the RACS frequency of 887~MHz. Our RACS result for a flux cut of 15 mJy, which agrees with NVSS, does not follow the trend predicted by the model, however the 150 mJy TGSS flux density cut made in \citet{Siewert2021} corresponds to a 40 mJy flux cut in RACS (assuming $\alpha=0.75$). Applying this flux density cut using the $N_{side}=32$ RACS map, the inferred dipole direction shifts by $\Delta\theta=8.8\degree$, and the dipole amplitude increases to $\mathcal{D}=(1.94\pm0.37)\times10^{-2}$. As such, our results cannot rule out the frequency dependence predicted by \citet{Siewert2021}, but the obtained results from WISE AGN \citep{Secrest2021,Secrest2022,Dam2022} provide a strong argument against it. Though results are consistent with literature, as with many works concerning the dipole, the correctness of the methods could stand some further examination, as do the results that we have obtained here.

\subsection{The Poisson solution}

\begin{figure*}
    \centering
    \includegraphics[width=0.48\textwidth]{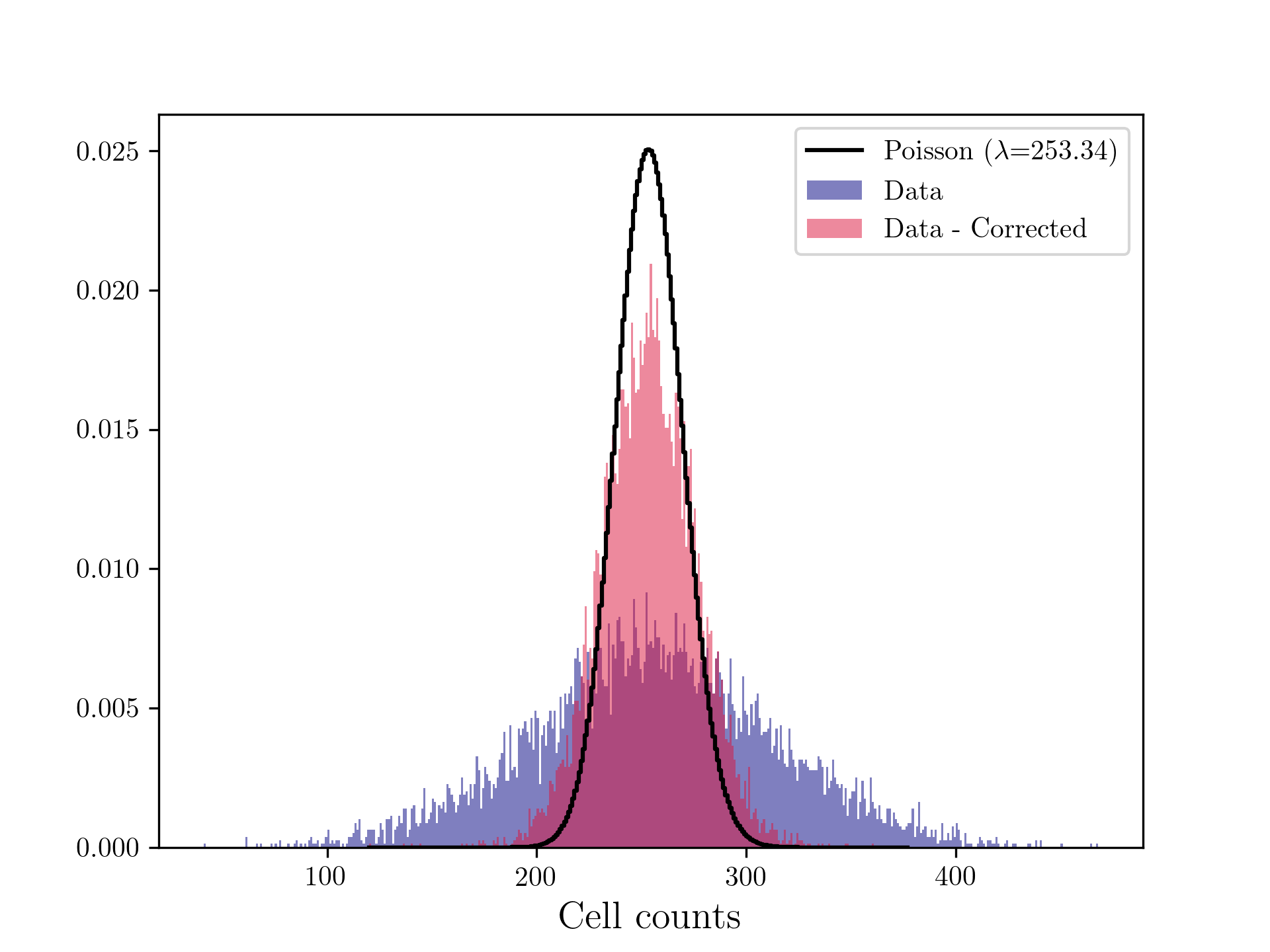}
    \includegraphics[width=0.48\textwidth]{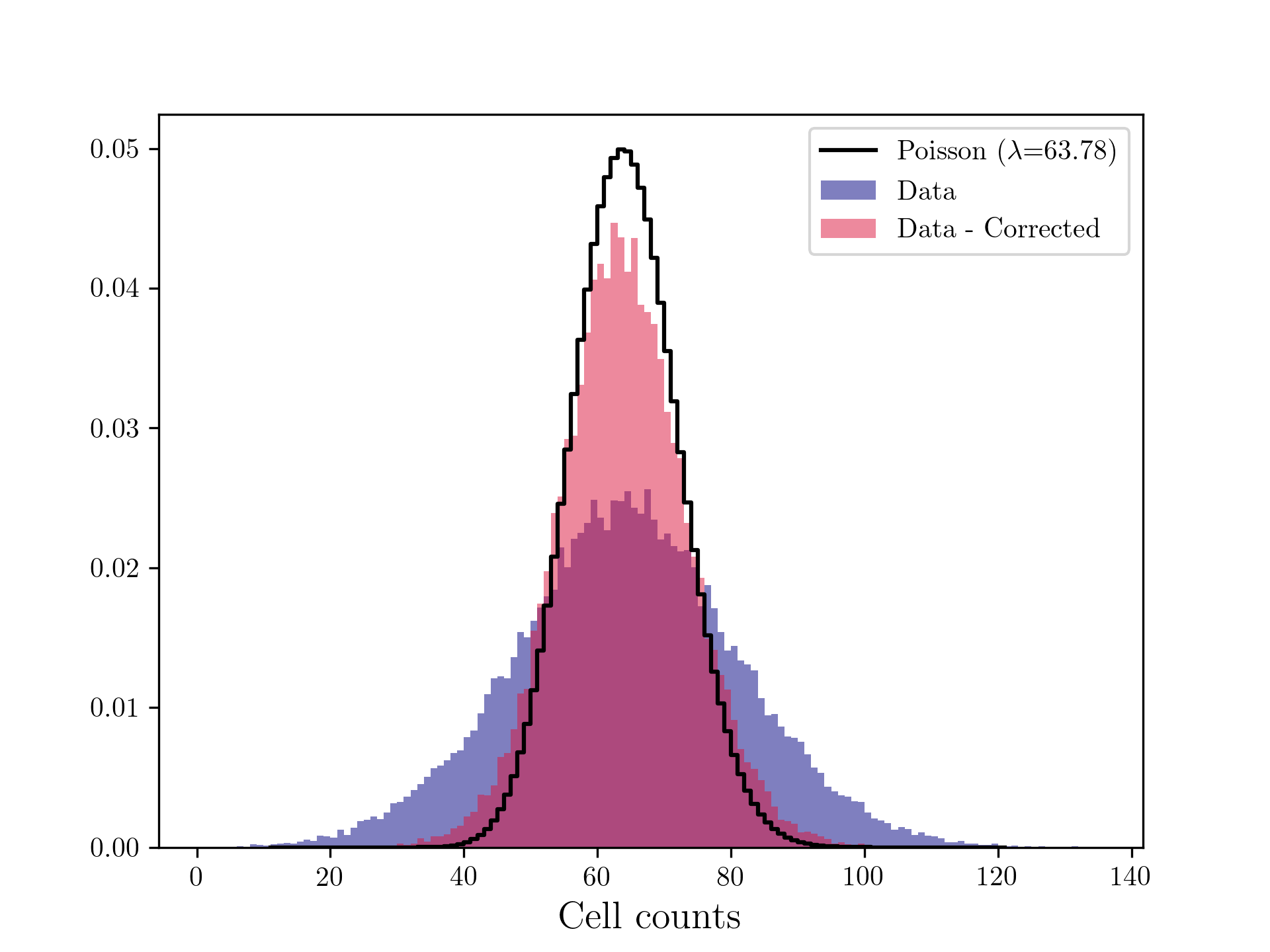}
    \caption{Cell count distribution for $N_{side}=32$ (left) and $N_{side}=64$ (right) RACS maps without any flux density cuts applied. Raw counts are shown (blue histogram) alongside the counts corrected for the power-law fit (red histogram). A Poisson distribution with a $\lambda$ equivalent to the estimated monopole amplitude is also shown (black histogram).}
    \label{fig:poisson_rms_racs}
\end{figure*}

Though we have shown here results that are both internally consistent and consistent with other dipole estimates, the choice of a Poisson estimator might seem like an unnecessary constraint on the data. After all, the quadratic estimator performs as adequately and does not suffer any loss in precision to the Poisson estimator. Table~\ref{tab:dipole_results} lists the $\chi^2$ values for the obtained results, defined by Equations~(\ref{eq:chi_square_quadratic}) and (\ref{eq:chi_square_poisson}) for the quadratic and Poisson estimators respectively. As the quadratic estimator is minimised for a Gaussian distribution with mean equal to the variance, it is expected that between the quadratic and basic Poisson estimators similar values are retrieved. This is indeed the case for the results in Table~\ref{tab:dipole_results}, both in the estimated parameters and $\chi^2$ values.

The value of the Poisson assumption becomes readily apparent when extending the parameter space, as we do when taking into account the RMS power-law relation. The main feature of a Poisson distribution is that one parameter is necessary to describe it, $\lambda$, which is both the mean and variance of the distribution. This is a strict requirement on a distribution, allowing more freedom in other parameters which would otherwise be degenerate with the parameters of the distribution. This means that fitting the RMS power law does not work with i.e. a quadratic estimator, which though minimised by a distribution with mean equal to the variance, still allows for a wider Gaussian distribution. As seen in Figure~\ref{fig:poisson_rms_racs}, the distribution of number counts without any flux density cut applied resembles a Gaussian distribution which is much wider than a Poisson distribution with the same mean. The quadratic estimator does however allow such a wide distribution, and thus will not converge on a solution that transforms this distribution to a Poisson distribution. Herein lies the power of the Poisson estimator, making modeling and fitting of systematics in the data a viable alternative to cutting and masking data. On the flip side of this, it is imposing a Poisson distribution on the data, which can lead to spurious results if improperly applied.

Table~\ref{tab:dipole_results} lists the $\chi^2/dof$ values of the Poisson RMS estimator after correction for the derived power law. The difference in distributions can be appreciated in Figure~\ref{fig:poisson_rms_racs}, which shows the distributions of the cell counts of RACS without any flux density cut applied, along with the same distribution corrected for the RMS power law that has been fit to the data. The uncorrected counts have a much wider distribution which is clearly not Poisson, with $\chi^2/dof=13.28$ for the $N_{side}=32$ map and $\chi^2/dof=4.49$ for the $N_{side}=64$ map. The corrected counts resemble a Poisson distribution more closely, with $\chi^2/dof=2.03$ for the $N_{side}=32$ map and $\chi^2/dof=1.49$ for the $N_{side}=64$ map, but are further from a Poisson distribution than the other results obtained, signifying that some residual effect has not been modelled by the estimator. 

As such, the performance of the Poisson-RMS estimator still leaves some questions to be answered. The assumption that source counts are related to sensitivity via a power law might carry a flaw, though there can be a number of possible reasons for this: (i) the median RMS is not the best representation of the sensitivity of the survey in a given cell, (ii) the sensitivity only properly represents source counts down to some limit, (iii) not all systematics equally impact source counts as well as sensitivity. These factors bear further examination in the future, but remarkable already are the results when comparing them to the other RACS results. It's clear that this Poisson-RMS estimator shows promise even in it's basic form, and can be used as an additional test of the data for any survey that has information on the local RMS. Furthermore, due to it's flexibility, additional effects once characterised can easily be modelled and taken into account by the estimator.

\subsection{Residual anisotropies in the data}

\begin{figure}
    \centering
    \includegraphics[width=\hsize]{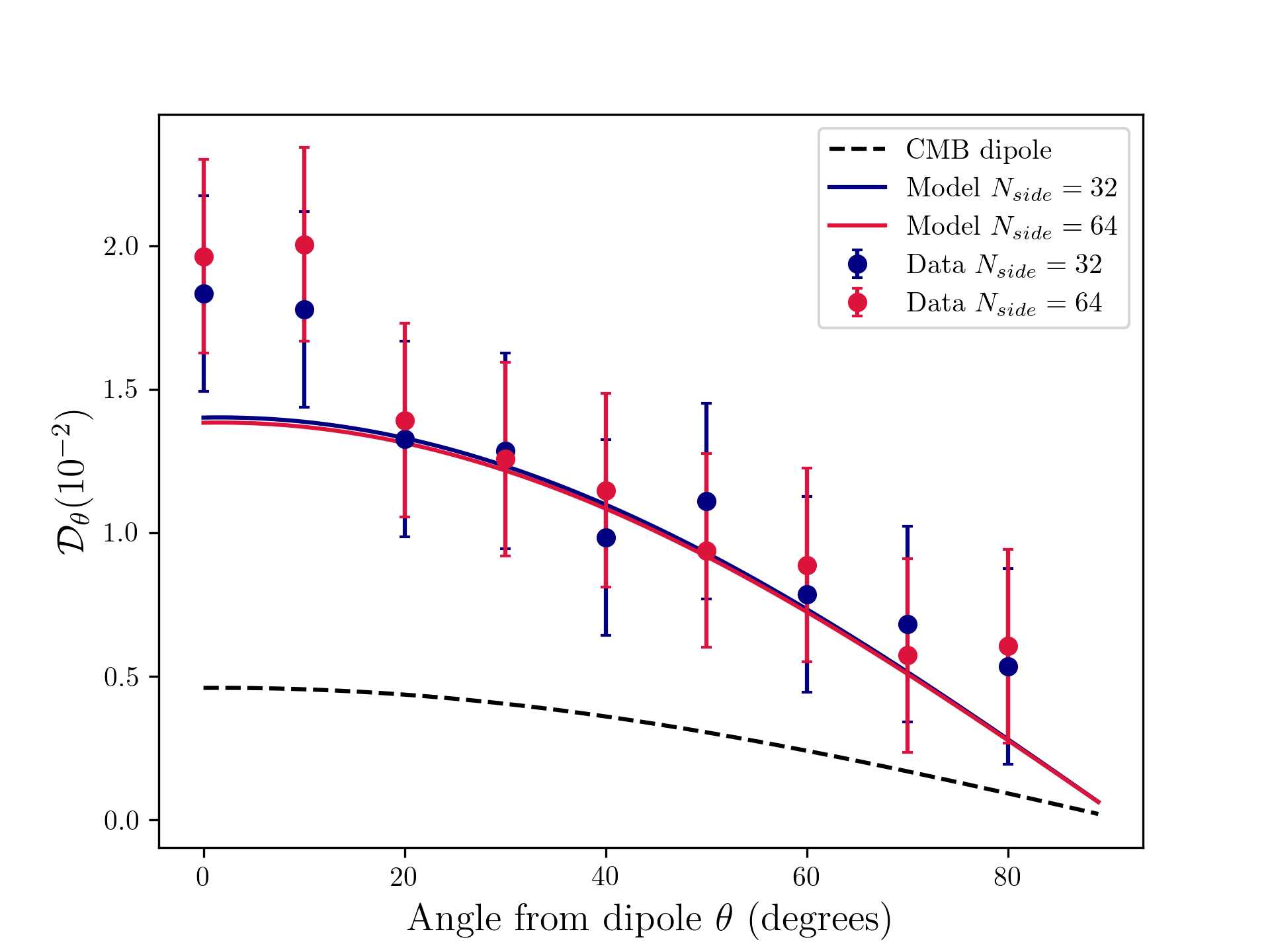}
    \caption{Dipole amplitude of NVSS as a function of angular distance to the dipole direction, assuming a best-fit dipole direction from the basic Poisson estimator, for both $N_{side}=32$ (blue) and $N_{side}=64$ (red) maps. Alongside the data the corresponding models are plotted (solid lines), as well as the expected model from the CMB dipole (black dashed line).}
    \label{fig:NVSS_hemisphere}
\end{figure}

\begin{figure}
    \centering
    \includegraphics[width=0.48\textwidth]{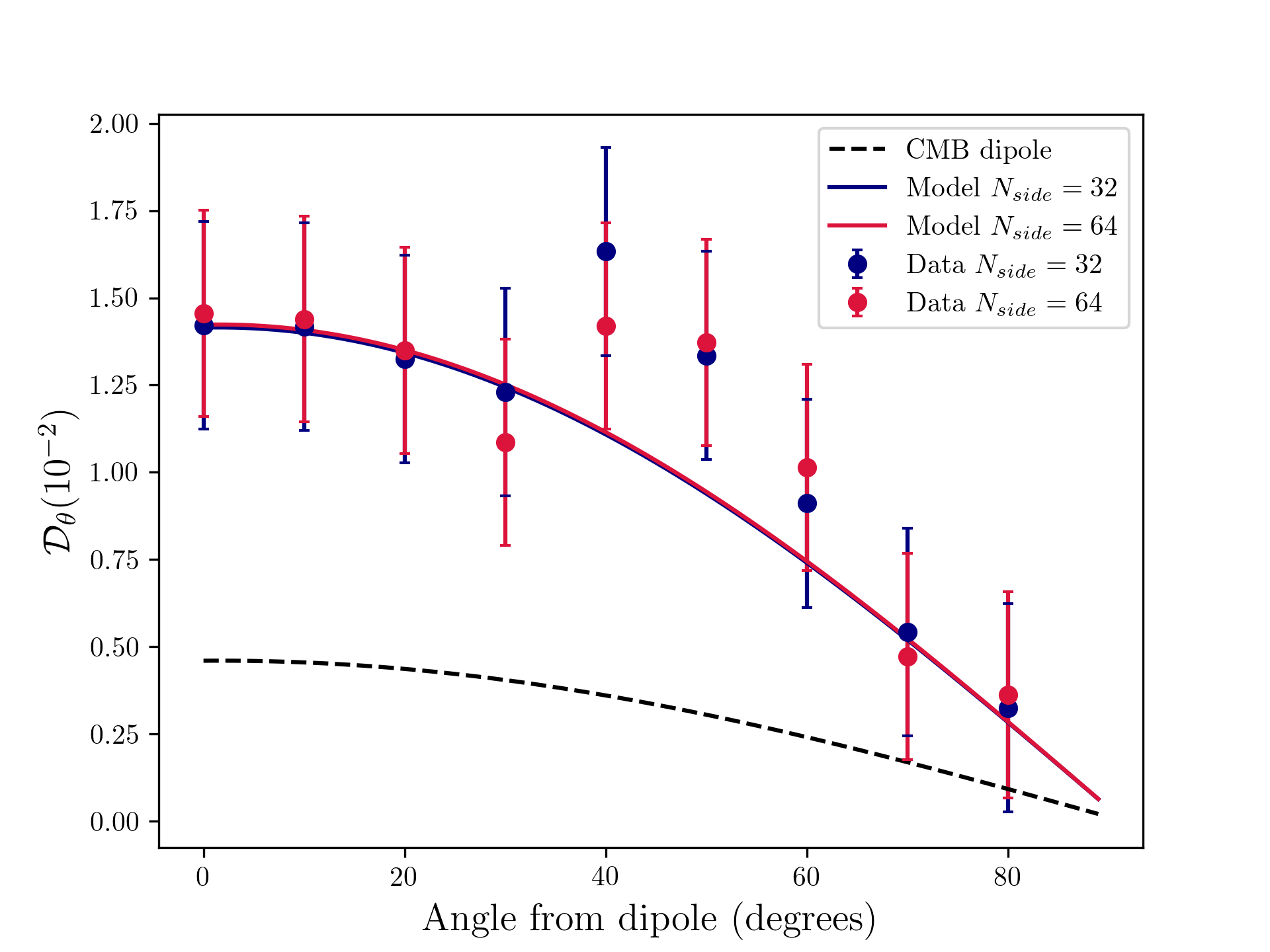}
    \includegraphics[width=0.48\textwidth]{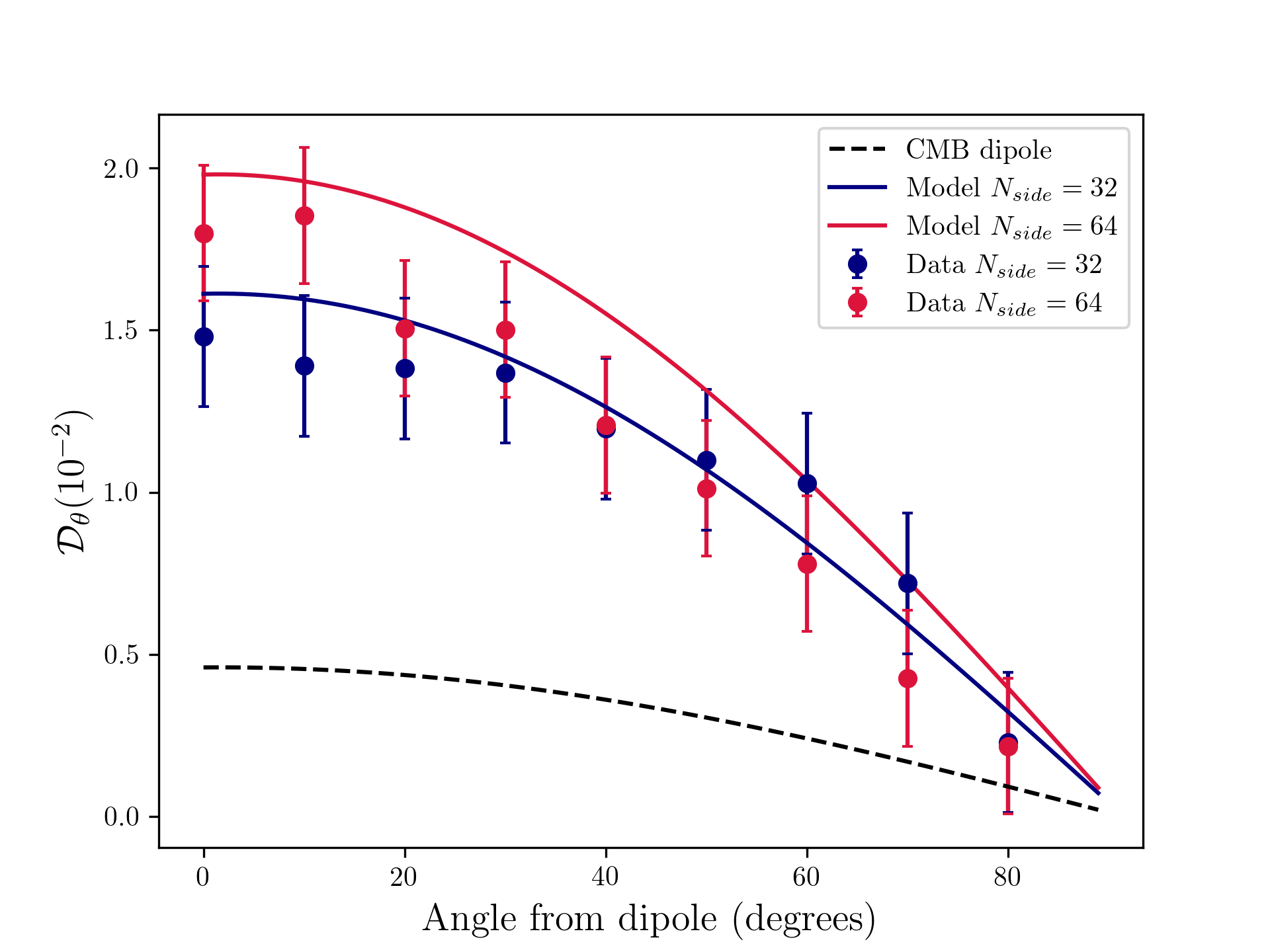}
    \caption{Dipole amplitude of RACS as a function of angular distance to the dipole direction, assuming a best-fit dipole direction from the basic Poisson estimator (top) and Poisson RMS estimator (bottom). Results for both $N_{side}=32$ (blue) and $N_{side}=64$ (red) maps are shown. Alongside the data the corresponding models are plotted (solid lines), as well as the expected model from the CMB dipole (black dashed line).}
    \label{fig:RACS_hemisphere}
\end{figure}

In dipole measurements and other statistical studies that require large amounts of data to retrieve a statistically significant measure, it can be difficult to visually assess whether any one fit adequately describes the data. After all we impose a model on the data to which the fit is restricted to. For a rudimentary visual check on whether the data follows the expected relations, we employ the hemisphere method used by \citet{Singal2021}. This method assumes that the direction of the dipole is already known, leaving as the only free parameter the dipole amplitude as a function of angular distance from the dipole direction, $\mathcal{D}_{\theta} = \mathcal{D}\cos\theta$. To reach statistically significant number counts, the sky is divided in two hemispheres, hemisphere $N_1$ with all sources between $\theta$ and $\theta + \pi/2$, and hemisphere $N_2$ with all sources between $\theta + \pi/2$ and $\theta + \pi$. The dipole amplitude as a function of $\theta$ is then written as
\begin{equation}
    \mathcal{D}_{\theta} = \frac{N_1(\theta) - N_2(\theta)}{\frac{1}{2}[N_1(\theta) + N_2(\theta)]}.
\end{equation}

We determine and plot the hemisphere results for NVSS and RACS assuming the results obtained from the Poisson estimators for the individual catalogues. The hemisphere relation for NVSS is shown in Figure~\ref{fig:NVSS_hemisphere} showing the data following the expected dipole curve except for the hemispheres closest to the dipole direction, which show an increased anisotropy. This is more pronounced in the $N_{side}=64$ map, where both $\theta=0\degree$ and $\theta=10\degree$ hemispheres show significantly increased counts compared to the expectation from the dipole model. In the $N_{side}=32$ map only the $\theta=0\degree$ hemisphere shows increased counts, with all other points following the dipole model within uncertainties. This points to a residual anisotropy left in the data that hasn't influenced the overall fit.

For RACS, the hemisphere relations for both the basic Poisson estimator as well as the RMS Poisson estimator are shown in Figure~\ref{fig:RACS_hemisphere}. It stands out immediately, especially in the results for the basic Poisson estimator, that there is a residual anisotropy in RACS at $40\degree - 60\degree$ from the dipole direction for both $N_{side}=32$ and $N_{side}=64$ maps. In case of the basic Poisson estimator, as with NVSS, the fit is unaffected. This anisotropy however might have had a significant impact on the RMS Poisson estimator for the $N_{side}=64$ map, as the dipole direction estimated from that map is $47\degree$ offset from the direction of the basic Poisson estimator, coinciding perfectly with the anisotropy seen at that angle. Indeed, the data for both $N_{side}=32$ and $N_{side}=64$ maps for the RMS Poisson estimate agree well with the exception of the points closest to the dipole direction, which in the case of the $N_{side}=64$ map are dominating the fit.

Finally, we investigate the possibility that residual systematics are present due to Galactic synchrotron. \citet{Secrest2022} use the de-striped and source-subtracted \citet{Haslam1982} 408 MHz all-sky map from \citet{Remazeilles2015} to mask pixels bright in Galactic synchrotron. To investigate if our results are impacted by Galactic synchrotron, we cross-correlate our number count maps with the \citet{Remazeilles2015} map. In all cases, no significant correlation is found ($|\rho| < 0.05$), showing that Galactic synchrotron is not present as a residual systematic in the data.

\subsection{Combining and splitting catalogues}

\begin{figure}
    \centering
    \includegraphics[width=\hsize]{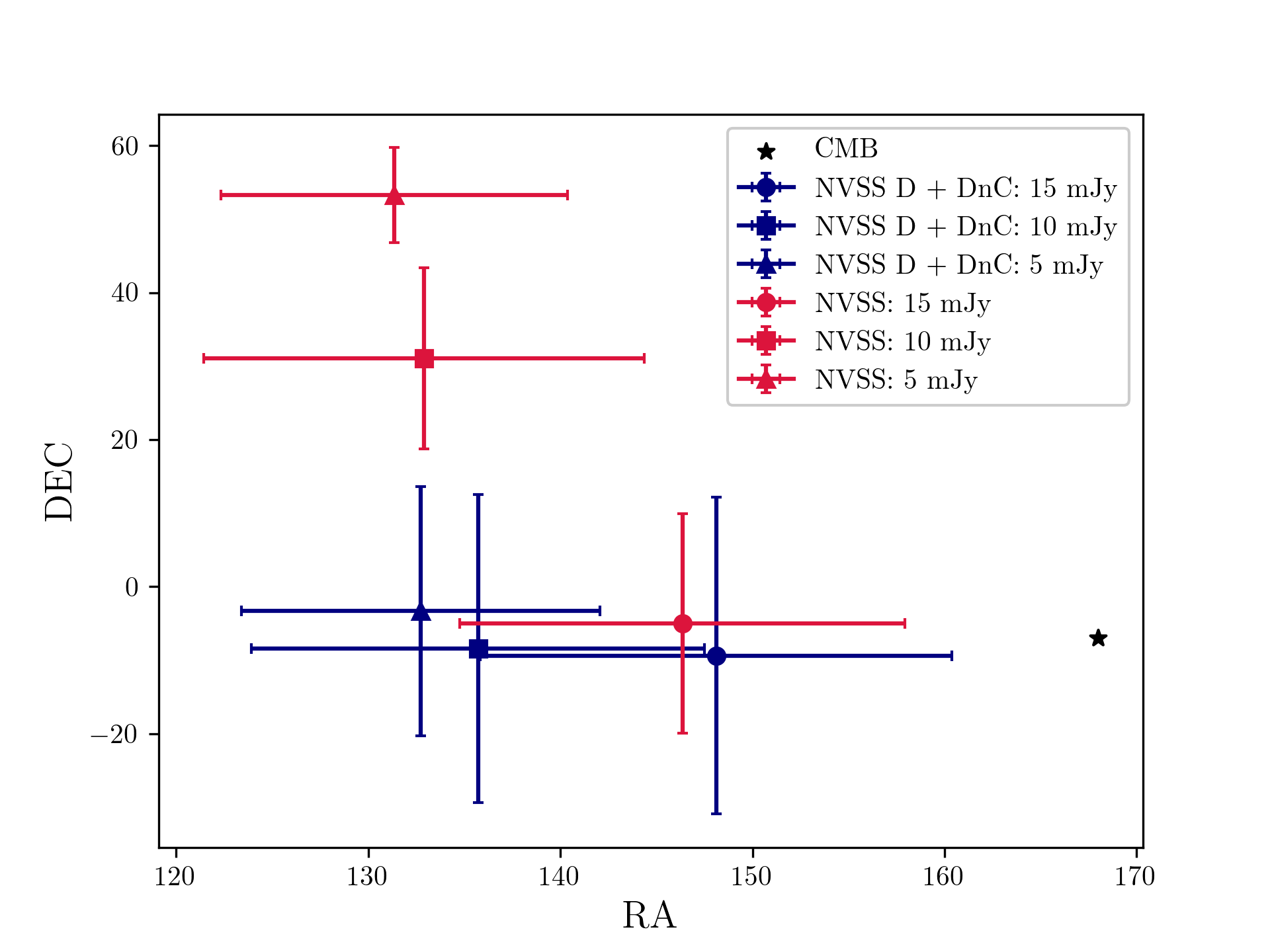}
    \caption{Best-fit dipole directions with $1\sigma$ uncertainties for the complete NVSS (red) and split NVSS D + DnC (blue) catalogues, for lower flux density thresholds of 5, 10, and 15 mJy. The CMB dipole direction is indicated with a black star.}
    \label{fig:nvvs_dirs}
\end{figure}

\begin{table*}[]
    \renewcommand*{\arraystretch}{1.4}
    \centering
    \caption{Dipole estimates for NVSS using different flux density cuts, separating D and DnC configurations.}
    \begin{tabular}{c | c c | c c | c c c}
    Catalogue &  $S_0$ & $N$ & $\mathcal{M}_1$ & $\mathcal{M}_2$ & $\mathcal{D}$ & R.A. & Dec. \\
     & (mJy) & & counts/pixel & & ($\times10^{-2}$) & (deg)  & (deg) \\
    \hline \hline
    NVSS & 15 & 345,803 & $40.4\pm0.1$ & -- & $1.40\pm0.29$ & $146\pm12$ & $-5\pm15$ \\
         & 10 & 480,446 & $56.1\pm0.1$ & -- & $1.38\pm0.26$ & $133\pm11$ &
    $31\pm12$ \\
         & 5  & 795,135 & $92.6\pm0.1$ & -- & $1.92\pm0.23$ & $131\pm9$ & $53\pm6$ \\
    \hline
    NVSS D + NVSS DnC & 15 & 333,046 & $40.5\pm0.1$ & $40.4\pm0.1$ & $1.44\pm0.31$ & $148\pm12$ & $-9\pm21$ \\
    & 10 & 463,398 & $56.4\pm0.1$ & $55.7\pm0.2$ & $1.23\pm0.27$ & $136\pm12$ & $-8_{-20}^{+22}$ \\
    & 5 & 767,832 & $93.5\pm0.2$ & $91.3\pm0.2$ & $1.18\pm0.20$ & $133\pm9$ & $-3_{-16}^{+18}$ 
    \end{tabular}
    \label{tab:nvss_results}
\end{table*}

As shown in Section~\ref{sec:combine}, combining catalogues as independent tracers of the dipole can provide a more robust measurement of the dipole, and in the case of NVSS and RACS, resulted in a dipole which matched the direction of the CMB dipole remarkably well, with a dipole amplitude 2.5 times the CMB expectation. The justification for this approach is that radio data is a complex product that is difficult enough to homogenise over a full survey internally, let alone between surveys. Field of view, frequency, array configuration, calibration, imaging and source finding are all factors to consider when assessing the source counts in a given survey. For a key example of how these factors can influence source counts, we need to look no further than NVSS, which has been observed with two different array configurations at certain declination ranges. This is a systematic that produces different source densities depending on the configuration, something that can be plainly seen in the left plot of Figure~\ref{fig:dec_source_density}.

To demonstrate the potential of the Multi-Poisson estimator beyond combining independent catalogues, we redo the dipole estimates for NVSS, lowering the minimum flux density in several steps. We then split the NVSS D and DnC configurations into separate catalogues, and repeat the experiment. The results are shown in Table~\ref{tab:nvss_results} and Figure~\ref{fig:nvvs_dirs}, showing the effect which splitting the configuration has on the dipole estimates. It stands out immediately that though the dipole estimate in both cases stays consistent in terms of right ascension, both in terms of dipole amplitude and declination of the dipole direction the separation of the catalogue produces wildly different results for flux density cuts below 15 mJy. This result can be more or less expected, as the difference in source density between the D and DnC configuration is expected to produce an anisotropy in the north-south direction, which is largely alleviated (although not entirely) with the split in configurations. Not only is this reflected in the declination of the dipole direction, but in the dipole amplitude as well, which is actually seen to decrease with more samples in the case of split configurations. Though the anisotropy in the declination is alleviated by this, another anisotropy in right ascension seems to start dominating at lower flux densities, dragging the dipole direction $43\degree$ from the CMB dipole direction. 

Though it seems that the different NVSS configurations produces an anisotropy that is on a similar level with other anisotropies related to incompleteness of the catalogue and cannot therefore be reliably used to completely account for the systematic effects in the survey that appear when employing lower flux density thresholds, these results show that were such systematic effects to dominate the catalogue, restructuring the problem to consider these as multiple independent catalogues can produce sensible results. It is furthermore a useful test of the chosen flux density threshold, as for an appropriately chosen flux density threshold the results between the full catalogue and split catalogue should be consistent. The results here thus show that a flux density threshold of 15 mJy is indeed appropriate for the NVSS. Such an approach might become very relevant given that VLASS also uses different array configurations, which can be taken into account when estimating a dipole in the same manner as we have done for NVSS. Even for RACS a prominent feature is a dependence of source counts on declination, though the exact mechanism is unclear, one possibility is related to the point-spread-function as the UV-coverage of the array evolves with declination.   

\subsection{Combining catalogues and cosmological considerations}

The combined dipole estimate of RACS and NVSS makes a compelling case to combine more probes of the cosmic dipole to increase sensitivity. The approach used here ostensibly carries less caveats than previous works, foregoing source matching, frequency scaling, sub-sampling, or weighting schemes that can all introduce additional uncertainties. This carries with it a reduction in formal uncertainty, though there remain some factors regarding the nature of the dipole that can limit the approach. As the nature of the excess amplitude of the radio dipole is currently not known, the approach of combining catalogues, even if done perfectly, carries an additional uncertainty. In Section~\ref{sec:data} the expected kinematic dipole amplitudes for both NVSS and RACS were computed and found to be nearly identical. This in large part justifies getting a combined estimate of the catalogues, however if we were to combine catalogues where the expected dipole amplitudes differed (e.g. when combining catalogues from multiple wavelengths), the approach cannot produce a reliable result without knowing the nature of the radio dipole. As it stands, we have a kinematic expectation of the radio dipole derived from the CMB. Given the velocity of the observer, this dipole is determined by the spectral index and flux distribution of the ensemble of sources. However, the excess dipole as it has been measured has an unknown origin, be it either kinetic or some entirely different effect. 

Given the results obtained so far in this work and the literature, one could also assume an effect which somehow boosts the observed dipole w.r.t. the CMB dipole, or an additional anisotropy which is simply added onto the kinematic dipole. Therefore should we wish to combine i.e. the sample of WISE AGN, which has an expected dipole amplitude of $\mathcal{D} = 7.3\times10^{-3}$ \citep{Secrest2022}, with for example NVSS, a combined estimate will have to assume one of these models. In fact, \citet{Secrest2022} find that after removing the CMB dipole, assuming it is purely kinematic, the residuals dipoles between NVSS and WISE agree with each other, favouring the interpretation of an intrinsic dipole anisotropy in the CMB rest frame. The results we have obtained for NVSS and RACS also support this interpretation, the residual dipole amplitudes being $\mathcal{D} = (0.97 \pm 0.30) \times 10^{-2}$ and $\mathcal{D} = (0.99 \pm 0.24) \times 10^{-2}$ respectively. The expected dipoles of the catalogues are however too similar to rule out other interpretations.

Furthermore, depending on which of the models presented above is true, survey design can have a profound impact on the measured cosmic radio dipole. The largest impact will be in the detected source populations and their redshift distributions. Naturally, going to optical or infrared wavelengths will yield different source populations that possibly trace the dipole differently, but even amongst radio surveys the detected source population will depend on the survey details. Radio surveys must be designed with a balance of depth and sky coverage, so we imagine a scenario where number of sources detected will stay constant over the survey due to this balance, thus not changing the significance of a dipole measurement. A shallow but large sky coverage radio survey will mostly detect AGN with a peaked redshift distribution, whereas a deep radio survey with limited sky coverage will probe most of the AGN population over all redshifts as well as star-forming galaxies, which have a different redshift distribution from AGN altogether. Most surveys which have been used for dipole estimates fall into the first category, as proper coverage along the dipole axis is necessary. However, a survey falling into the second category would have the potential to differentiate between the possible models of the radio dipole  we have laid out. The MeerKAT Absorption Line Survey \citep{Gupta2016}, consisting of sparsely spaced deep pointings homogeneously distributed across the sky, provides a good candidate for a survey falling into this second category, and thus might provide more insight behind the processes driving the anomalous amplitude of the radio dipole. 

\section{Conclusion}
\label{sec:conclusion}

In this work we have presented a set of novel Bayesian estimators for the purpose of measuring the cosmic radio dipole  with the NVSS and RACS catalogues. Based on the assumption that counts-in-cell of radio sources follow a Poisson distribution, we construct estimators for the cosmic radio dipole  based on Poisson statistics. As a basis of comparison we include a quadratic estimator, which has been used in a number of previous dipole studies. We furthermore construct two extensions of the basic Poisson estimator to attempt to account for systematic effects in the respective catalogues. Firstly, we consider that if sensitivity information is present in the catalogue, this can be directly related to the local number density assuming that systematic effects merely modify the local sensitivity of the catalogue. The local sensitivity and number counts are assumed to be related by power law, the parameters of which can be estimated. We extend the Poisson estimator to address this. Secondly, we construct an extension to the Poisson estimator that can be given multiple separate catalogues, assuming that the catalogues trace the same dipole, and produces a combined estimate.

We obtain best-fit parameters for the cosmic radio dipole  by using $\chi^2$ minimisation for the quadratic estimator and using maximum likelihood estimation for the Poisson estimators. To discretise the sky, we use \textsc{HEALPix}, producing maps with both $N_{side}=32$ and $N_{side}=64$. We verify that the quadratic estimator and basic Poisson estimator yield similar results, and that furthermore results between the pixel scales are consistent. We use the Poisson RMS estimator on RACS while not using any cut in flux density to estimate the dipole parameters along with the parameters for the RMS power law. The increased number counts greatly increase precision of the estimate, but the results somewhat diverge from the dipole estimates produced by the basic estimators. Whether this difference is a genuine product of the data or a flaw in (assumptions of) the estimator cannot be said for certain, however the initial results are still promising given that the entire catalogue of sources is used. We finally use Poisson estimator for multiple catalogues on NVSS and RACS and obtain a dipole estimate that perfectly aligns with the CMB dipole in terms of direction, but has an amplitude three times as large with a significance of $4.8\sigma$. Given the dipole estimates obtained from the individual catalogues, this result is in line of expectations for a combination of the two catalogues, and thus can be seen as the most reliable and significant result obtained here.

We explore the possibility of splitting up a catalogue and using the Poisson estimator for multiple catalogues to estimate the dipole as if on two independent catalogues. We use this method on the NVSS, which has been observed with two different array configurations, which introduces and artifical north-south anisotropy in the catalogue. We treat these array configurations as separate catalogues and repeat the dipole estimate, and go down to lower flux density limits than with the basic Poisson estimator. We see that while using the whole NVSS a north-south anisotropy starts dominating the estimate at lower flux densities, separating the configuration largely mitigates this effect. As a result, this allows us to lower the flux density cut, increasing number counts and thus increasing the significance of the dipole estimates. This approach may serve well on catalogues such as VLASS, which also uses different array configurations in different parts of the sky. Through the presented estimator there has potential in combining a larger variety of catalogues, the extent to which this can be done depends in large part on the nature of the excess dipole. With ever more probes at various wavelengths, sky coverage, and depth reaching the necessary sensitivity to detect the dipole, these might shed light on the nature of this dipole, given the different populations of sources that are probed. 

\begin{acknowledgements}
We thank the anonymous referee for their useful comments and feedback on the text.
JDW acknowledges the support from the International Max Planck Research School (IMPRS) for Astronomy and Astrophysics at the Universities of Bonn and Cologne.
Some of the results in this paper have been derived using the healpy \citep{Zonca2019} and \textsc{HEALPix} \citep{Gorski2005} packages.
This research has made use of \textsc{topcat} \citep{Taylor2005}, \textsc{lmfit} \citep{Newville2016}, \textsc{bilby} \citep{Ashton2019}, and \textsc{emcee} \citep{Foreman-Mackey2013}.
\end{acknowledgements}

\bibliographystyle{aa}
\bibliography{main_paper}

\end{document}